\documentclass[useAMS,usenatbib,usegraphicx]{mn2e}

\def\ltsima{$\; \buildrel < \over \sim \;$}
\def\simlt{\lower.5ex\hbox{\ltsima}}
\def\gtsima{$\; \buildrel > \over \sim \;$}
\def\simgt{\lower.5ex\hbox{\gtsima}}
\def\gsimeq
{\hbox{\raise0.5ex\hbox{$>\lower1.06ex\hbox{$\kern-1.07em{\sim}$}$}}}
\def\lsimeq
{\hbox{\raise0.5ex\hbox{$<\lower1.06ex\hbox{$\kern-1.07em{\sim}$}$}}}

\def\xmm{{\it XMM-Newton }}

\def\xmm{{\it XMM-Newton}}
\def\chandra{{\it Chandra}}

\def\nustar{{\it NuSTAR}}
\def\sinfoni{{\it SINFONI}}

\def\apj{ApJ}
\def\aj{AJ}
\def\mnras{MNRAS}
\def\aap{A\&A}

\def\apjl{ApJ}
\def\apjs{ApJS}
\def\araa{ARA\&A}

\def\ssr{SSRv}
\def\nat{Nature}

\def\procspie{Proc. SPIE}

\def\sgr{Sgr~A$^{\star}$}

\def\axj{AX~J1745.6-2901}
\def\sgr{SGR J1745-2900}
\def\ssv{Swift~J174540.7-290015}

\def\sgras{Sgr~A$^\star$}

\def\xis{XIS}
\def\xis1{XIS1}
\def\xis2{XIS2}
\def\xis3{XIS3}

\title[] 
 {{A powerful flare from \sgras\ confirms the synchrotron nature of 
 the X-ray emission}}

 \author[G.\ Ponti et al. ]
 {G.~Ponti$^{1}$\thanks{ponti@mpe.mpg.de}, 
 E.~George$^{1}$, 
 S.~Scaringi$^{1,2}$, 
 S.~Zhang$^{3,4}$, 
 C.~Jin$^{1}$,
 J. Dexter$^{1}$,
 R. Terrier$^{5,6}$,
  \newauthor
 M. Clavel$^{7}$,
 N. Degenaar$^{8,9}$,
 F. Eisenhauer$^{1}$,
 R. Genzel$^{1}$, 
 S. Gillessen$^{1}$,
 A. Goldwurm$^{5,6}$,
   \newauthor
 M. Habibi$^{1}$,
 D. Haggard$^{10,11}$,
 C. Hailey$^{2}$,
 F. Harrison$^{12}$,
 A. Merloni$^{1}$,
 K. Mori$^{2}$,
    \newauthor
 K. Nandra$^{1}$,
 T. Ott$^{1}$ 
 O. Pfuhl$^{1}$,
 P. M. Plewa$^{1}$,
 and I. Waisberg$^{1}$ \\
   $^1$ Max Planck Institute fur Extraterrestriche Physik, 85748, Garching, Germany\\
   $^2$ Department of Physics and Astronomy, University of Canterbury, Private Bag 4800 Christchurch, New Zealand \\
   $^3$ Columbia Astrophysics Laboratory, Columbia University, New York, NY 10027, USA \\
   $^4$ MIT Kavli Institute for Astrophysics and Space Research \\
   $^5$ Unit\'e mixte de recherche Astroparticule et Cosmologie, 10 rue Alice Domon 
   et L\'eonie Duquet, 75205 Paris, France \\
   $^6$ Service d'Astrophysique (SAp), IRFU/DRF/CEA-Saclay, 91191 
   Gif-sur-Yvette Cedex, France\\
   $^7$ Space Sciences Laboratory, 7 Gauss Way, University of California, Berkeley, CA 94720-7450, USA \\
   $^{8}$ Institute of Astronomy, University of Cambridge, Madingley Road, Cambridge CB3 OHA, UK \\
   $^{9}$ Anton Pannekoek Institute for Astronomy, University of Amsterdam, Science Park 904, 1098 XH, Amsterdam, the Netherlands \\
   $^{10}$ McGill University, Department of Physics, 3600 rue University, Montr\'eal, QC, H3A 2T8 \\
   $^{11}$ McGill Space Institute, 3550 rue University, Montr\'eal, QC, H3A 2A7 McGill \\
   $^{12}$ Jet Propulsion Laboratory, California Institute of Technology, 4800 Oak Grove Drive, Mail Stop 169-221, Pasadena, CA 91109, USA \\
}

\pagerange{\pageref{firstpage}--\pageref{lastpage}}

\usepackage{arydshln}
\usepackage{times}

\begin{document}

\label{firstpage}

\maketitle

\begin{abstract}
We present the first fully simultaneous fits to the NIR and X-ray spectral 
slope (and its evolution) during a very bright flare from \sgras, the 
supermassive black hole at the Milky Way's center. Our study arises 
from ambitious multi-wavelength monitoring campaigns with \xmm, 
\nustar\ and \sinfoni. 
The average multi-wavelength spectrum is well reproduced by a broken 
power-law with $\Gamma_{NIR}=1.7\pm0.1$ and $\Gamma_X=2.27\pm0.12$.
The difference in spectral slopes ($\Delta\Gamma=0.57\pm0.09$) strongly 
supports synchrotron emission with a cooling break. 
The flare starts first in the NIR with a flat and bright NIR spectrum, 
while X-ray radiation is detected only after about $10^3$~s, when a very 
steep X-ray spectrum ($\Delta\Gamma=1.8\pm0.4$) is observed. 
These measurements are consistent with synchrotron emission 
with a cooling break and they suggest that the high energy cut-off in 
the electron distribution ($\gamma_{max}$) induces  an initial cut-off 
in the optical-UV band that evolves slowly into the X-ray band. 
The temporal and spectral evolution observed in all bright X-ray 
flares are also in line with a slow evolution of $\gamma_{max}$. 
We also observe hints for a variation of the cooling break that 
might be induced by an evolution of the magnetic field (from 
$B\sim30\pm8$~G to $B\sim4.8\pm1.7$~G at the X-ray peak). 
Such drop of the magnetic field at the flare peak would be expected if 
the acceleration mechanism is tapping energy from the magnetic field, 
such as in magnetic reconnection. We conclude that synchrotron emission with a cooling 
break is a viable process for \sgras's flaring emission. 
\end{abstract}


\normalsize
\section{Introduction} 
\label{intro}

\sgras, the supermassive black hole (BH) at the Milky Way's center, with 
a bolometric luminosity of $L\sim10^{36}$~erg~s$^{-1}$ is currently characterised 
by an exceptionally low Eddington ratio ($\sim10^{-8}$; Genzel et al. 2010a), 
despite indications that \sgras\ might have been brighter 
in the past (see Ponti et al. 2013 for a review). 
Therefore, \sgras\ provides us with the best chance to get a glimpse of the 
physical processes at work in quiescent BH. 

\begin{figure}
\includegraphics[height=0.49\textwidth,angle=90]{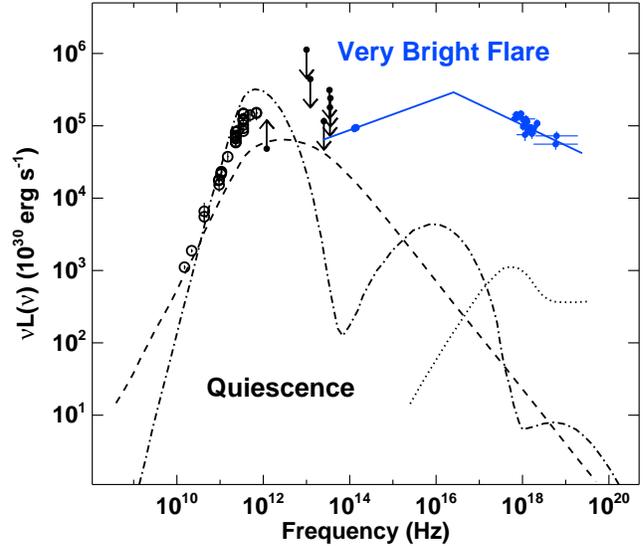}
\caption{Multi-wavelength emission from \sgras. The radio to mid-IR data 
points (open circles as well as upper-lower limits) constrain \sgras's quiescent 
emission (these constraints are taken from the literature, see text). The black 
lines show the radiatively inefficient accretion disc model proposed by Yuan 
et al. (2003). The dash-dotted black line shows the contribution from the thermal 
electrons with $T_e\sim few~10^{10}~K$ and $n_e\sim10^6$~cm$^{-3}$, 
embedded in a magnetic field with strength of $\sim10-50$~G producing the 
sub-mm peak and (possibly) inverse Compton emission at higher energies. 
The dashed line shows the contribution from a non-thermal tail in the electron 
population, while the dotted line shows the bremsstrahlung 
emission from hot plasma at the Bondi radius (Quataert 2002).
The blue filled data points shows the mean NIR and X-ray spectra of the very 
bright flare VB3 and the blue solid line shows the best fit power-law 
with cooling break (see also Fig. 6). }
\label{SED}
\end{figure}

\sgras\ has been intensively studied over the past several decades 
at various wavelengths. The black points 
(upper-lower limits) in Fig. \ref{SED} show a compilation 
of measurements of \sgras's quiescent emission from radio to mid-IR 
(values are taken from Falcke et al. 1998; Markoff et al. 2001; An et al. 2005; 
Marrone et al. 2006; Sch{\"o}del et al. 2007; 2011; Dodds-Eden et al. 2009; 
Bower et al. 2015; Brinkerink et al. 2015; Liu et al. 2016; Stone et al. 2016) 
as well as the radiatively inefficient accretion flow model proposed 
by Yuan et al. (2003). The bulk of \sgras's steady 
radiation is emitted at sub-mm frequencies, forming the so called "sub-mm 
bump" (dot-dashed line in Fig. \ref{SED}). This emission is linearly polarised 
(2-9~\%; Marrone et al. 2006; 2007), slowly variable and decreases rapidly 
with frequency (with stringent upper limits in the mid-IR band; Sch\"{o}del et 
al. 2007; Trap et al. 2011). This indicates that the sub-mm radiation is 
primarily due to optically thick synchrotron radiation originating in the central 
$\sim10$~$R_S$\footnote{$R_S$ is the Schwarzschild radius 
$R_S=2GM_{BH}/c^2$, where $M_{BH}$ is the BH mass, $G$ the 
gravitational constant and $c$ the speed of light.} and produced by 
relativistic ($\gamma_e\sim10$, where $\gamma_e$ is the electron 
Lorentz factor) thermal electrons with temperature 
and densities of $T_e\sim$ few $10^{10}~K$ and $n_e\sim10^6$~cm$^{-3}$, 
embedded in a magnetic field with a strength of $\sim10-50$~G (Loeb \& Waxman 
2007; Genzel et al. 2010a; in Fig. \ref{SED} a possible inverse Compton 
component is also shown). Moreover, Faraday rotation measurements 
constraint the accretion rate at those scales to be within 
$2\times10^{-9}$ and $2\times10^{-7}$~M$_\odot$ yr$^{-1}$ (Marrone 
et al. 2006; 2007; Genzel et al. 2010a). 

At low frequency ($\nu<10^{11}$~Hz) \sgras's SED changes slope 
($F_\nu\propto\nu^{0.2}$) showing excess emission above 
the extrapolation of the thermal synchrotron radiation and variability 
on time-scales of hours to years (Falcke et al. 1998; Zhao et al. 2003; 2004;
Herrnstein et al. 2004). This suggests either the presence of a non-thermal tail 
in the electron population, taking $\sim1$~\% of the steady state electron 
energy (\"{O}zel et al. 2000; see dashed line in Fig. \ref{SED}) or a compact 
radio jet (Falcke et al. 1998; Mo\'scibrodzka et al. 2009; 2013; 2014).
The presence of this non-thermal tail is well constrained at low radio 
frequencies, while its extrapolation in the mid and near infra-red band 
is rather uncertain (see dashed line in Fig. \ref{SED}). 

\sgras\ also appears as a faint ($L_{2-10~keV}\sim2\times10^{33}$~erg~s$^{-1}$) 
X-ray source (Baganoff et al. 2003; Xu et al. 2006) observed to be extended 
with a size of about $\sim1^{\prime\prime}$. The observed size is comparable 
to the Bondi radius and the quiescent X-ray emission is thought to be the 
consequence of material that is captured at a rate of $10^{-6}$~M$_\odot$ yr$^{-1}$ 
from the wind of nearby stars (Melia 1992; Quataert 2002; Cuadra et al. 2005; 
2006; 2008). Indeed, this emission is thought to be produced via 
bremsstrahlung emission from a hot plasma with $T\sim7\times10^7$~K, density 
$n_e\sim100$~cm$^{-3}$ emitted from a region $\sim10^5$~R$_S$ 
(Quataert 2002; see dotted line in Fig. \ref{SED}). 

For more than a decade it has been known that \sgras\ also shows flaring 
activity both in X-rays and IR (Baganoff et al. 2001; Goldwurm et al. 2003; 
Genzel et al. 2003; Ghez et al. 2004; Porquet et al. 2003; 2008; 
B\'elanger et al. 2005; Eckart et al. 2004; 2006; Marrone et al. 2008; 
Nowak et al. 2012; Haubois et al. 2012; Neilsen et al. 2013; 2015; 
Degenaar et al. 2013; Barri\`ere et al. 2014; Moussoux et al. 2015; 
Ponti et al. 2015a; Yuan \& Wang 2016). 
X-ray flares appear as clear enhancements above the 
constant quiescent emission, with peak luminosities occasionally exceeding 
the quiescent luminosity by up to two orders of magnitude 
(see the blue points in Fig. \ref{SED} for an example of a very bright flare; 
Baganoff et al. 2001; Porquet et al. 2003; 2008; Nowak et al. 2012). 
X-ray flare durations, fluences, and peak luminosities are correlated 
(Neilsen et al. 2013). Moreover, weak X-ray flares are more common 
than strong ones (Neilsen et al. 2013; Ponti et al. 2015a). 
The appearance of the X-ray light curves suggests that flares are individual 
and distinct events, randomly punctuating an otherwise quiescent source 
(Neilsen et al. 2013; Ponti et al. 2015a). 

Typically X-ray flares coincide with clear peaks in the near infrared (NIR) light 
curves (e.g. Genzel et al. 2003; Clenet et al. 2004; Ghez et al. 2004; 
Eckart et al. 2006; 2008; Meyer et al. 2006; 2007; 2008; 
Yusef-Zadeh et al. 2006; 2009; 
Hornstein et al. 2007; Do et al. 2009). However, the appearance of the NIR 
light curves is significantly different from the X-ray ones. Indeed, the NIR and 
sub-mm emission of \sgras\ are continuously varying and they can be described 
by a red noise process at high frequencies, breaking at time-scales longer 
than a fraction of a day\footnote{Interestingly, AGN of similar BH mass, 
but clearly much higher accretion rate, show power density spectra (PDS) 
of their X-ray light curves consistent with the NIR PDS of \sgras\ 
(Meyer et al. 2009). } (Do et al. 2009; Meyer et al. 2009; 
Dodds-Eden et al. 2011; Witzel et al. 2012; Dexter et al. 2014; Hora et al. 2014). 
Therefore, the NIR light curves do not support the notion of flares
as individual events; they would alternatively corroborate the concept that flares
are simply peaks of emission on a continuous red noise process. 
Despite the fact that the notion of NIR-flares is still unsettled, {\it we 
will refer to the X-ray flares, and by extension, to the NIR peaks in emission, 
as flares throughout this paper}.

The origin and radiative mechanism of the flares of \sgras\ are still 
not completely understood. Several multi-wavelength campaigns have been 
performed, but the radiative mechanisms at work during the flares is still highly 
debated (Eckart et al. 2004, 2006, 2008, 2009, 2012; Yusef-Zadeh et al. 2006, 2008, 
2009; Hornstein et al. 2007; Marrone et al. 2008; Dodds-Eden et al. 2009; 
Trap et al. 2011; Barriere et al. 2014). 
Indeed, even though 15 years have passed since 
the launch of \xmm\ and \chandra, simultaneous X-ray and NIR spectra 
of a bright flare (as these allow a precise determination of the spectral index 
in the two bands) has not yet been published\footnote{Indeed, the few 
multi-wavelength campaigns that caught a very bright X-ray flare were 
missing spectroscopic information in the NIR band (e.g., Dodds-Eden et 
al.\ 2009), while the few campaigns with spectroscopic information in 
both bands, failed to detect a bright X-ray flare with a NIR counterpart 
(e.g., flare A and B of Trap et al.\ 2011). }. 

Polarisation in the sub-mm and NIR bands suggests that the NIR radiation 
is produced by synchrotron emission. The origin of the X-ray emission 
is still debated. Indeed, the X-ray radiation could be 
produced by synchrotron itself or inverse Compton processes such as  
synchrotron self-Compton or external Compton (see Genzel et al. 2010a 
for a review). 
Different models explain the data with a large range of physical parameters, 
however, models with synchrotron emission extending with a break 
from NIR to the X-ray seem to be best able to account for the X-ray 
data with reasonable physical parameters (Dodds-Eden et al. 2009; 
Trap et al. 2010; Dibi et al. 2014; 2016; Barriere et al. 2014). 

We report here the first simultaneous observation of the X-ray (\xmm), 
hard X-ray (\nustar), and NIR (\sinfoni) spectra of a very bright flare of 
\sgras, which occurred between 2014 August $30^{th}$ and $31^{st}$
(Ponti et al. 2015a; 2015c), and an analysis of flare models which could explain 
the emission. The remainder of this paper is organised as follows:
Section \ref{datared} details the reduction of the X-ray and NIR data. 
In section \ref{brightxmm} we present a characterisation of the 
obscuration and mean spectral properties of the very bright flares 
observed by \xmm. 
In section \ref{multiwave} we investigate the mean properties 
of the VB3 flare, in particular we constrain the radiative mechanism 
through the study of the mean multi-wavelength spectrum. 
In section \ref{SecSpEvol}, we follow the evolution of the flare 
emission to determine time-dependent parameters of 
the emission models. Section \ref{quie} scrutinises a "quiescent" 
interval after the very bright flare. 
In Section \ref{SpEvol} we focus the analysis on the X-ray band 
only and we study the evolution of the X-ray spectral shape 
throughout all bright and very bright X-ray flares. 
We discuss the results of the model fits in section \ref{discussion} 
and conclude in section \ref{conclusions}.

\section{Data reduction} 
\label{datared} 

We consider two sets of data in this paper. 
The first set comprises simultaneous X-ray (\xmm\ and \nustar) and 
NIR data of one very bright flare, called VB3 (see Ponti et al. 2015 for the 
definition of the naming scheme). The analyses of the 
\xmm, \nustar\ and \sinfoni\ data on flare VB3 are discussed 
in sections \ref{sub:xmm}, \ref{sub:nustar} and \ref{sub:sinfoni}.
The second set of data consists of all of \sgras's bright or very bright 
X-ray flares as detected with \xmm. The reduction of this set of data 
is discussed in section \ref{sub:xmm} along with the description of 
the flare VB3. 

\subsection{Basic assumptions}
\label{Assumptions}

Throughout the paper we assume a distance to \sgras\ of 8.2~kpc 
and a mass of M$ = 4.4\times10^6$~M$_{\odot}$ (Genzel et al. 2010a). 
The errors and upper limits quoted on spectral fit results correspond 
to 90 \% confidence level for the derived parameters (unless 
otherwise specified), while uncertainties associated with measurements 
reported in plots are displayed at the $1~\sigma$ confidence level. 
The neutral absorption affecting the X-ray spectra is fitted with the model 
{\sc TBnew}\footnote{http://pulsar.sternwarte.uni-erlangen.de/wilms/research/tbabs/}  
(see Wilms et al. 2000) with the cross sections of Verner et al. (1996) and 
abundances of Wilms et al. (2000).
The dust scattering halo is fitted with the model {\sc fgcdust} in 
{\sc XSpec} (Jin et al. 2016; see \S \ref{brightxmm}) and 
it is assumed to be the same as the "foreground" component along 
the line of sight towards \axj\ (Jin et al. 2016). More details on the 
implications of this assumptions are included in \S \ref{brightxmm} 
and Appendix \ref{SecDSH}. 
In \S \ref{brightxmm} we justify the assumption of a column density 
of neutral absorbing material of $N_H=1.60\times10^{23}$~cm$^{-2}$
and we apply it consistently thereafter. Throughout our discussion 
we assume that the effects of beaming are negligible, as well as 
a single zone emitting model for the source. 

Unless otherwise stated, we follow Dodds-Eden et al. (2009) 
and we assume a constant escape time of the synchrotron 
emitting electrons equal to $t_{esc}=300$~s. Under this 
assumption, the frequency of the synchrotron cooling break 
can be used to derive the amplitude of the source magnetic field. 

\subsection{\xmm}
\label{sub:xmm}

In this work, we considered all of the \xmm\ observations during which 
either a bright or very bright flare has been detected through the Bayesian 
block analysis performed by Ponti et al. (2015a). 
Full details about the observation identification (obsID) are reported 
in Tab. \ref{obsid}. 

Starting from the \xmm\ observation data files, we reprocessed all of the data sets 
with the latest version (15.0.0) of the \xmm\ Science Analysis System (SAS), 
applying the most recent (as of 2016 April 27, valid for the observing day) calibrations. 
Whenever present, we eliminated strong soft proton background flares, 
typically occurring at the start or end of an observation, by cutting the exposure time 
as done in Ponti et al. (2015a; see Tab. 7). 
To compare data taken from different satellites and from the ground, we performed 
barycentric correction by applying the {\sc barycen} task of {\sc sas}. 
The errors quoted on the analysis of the light curves correspond 
to the 1 $\sigma$ confidence level (unless otherwise specified). 
{\sc XSpec} v12.8.2 and {\sc matlab} are used for the spectral analysis
and the determination of the uncertainties on the model parameters. 

\begin{table} 
\begin{center}
\small
\begin{tabular}{ c c c c c c }
\hline
\hline
\xmm\            & Date &t$_{\rm start}$   &t$_{\rm stop}$    & NAME \\
obsID             &         &TT$_{\rm TBD}$&TT$_{\rm TBD}$& \\
\hline
\multicolumn{5}{l}{\it Archival data} \\
\hdashline
0111350301 & 2002-10-03 &  150026757 &  150029675 &   {\bf VB1} \\
                     &                     &  150026757 &  150027730 &   VB1-Rise \\
                     &                     &  150027730 &  150028702 &   VB1-Peak \\
                     &                     &  150028702 &  150029675 &   VB1-Dec \\
\hdashline
0402430401 & 2007-04-04 &  292050970 &  292054635 &   {\bf VB2} \\
                     &                     &  292050970 &  292052192 &   VB2-Rise \\
                     &                     &  292052192 &  292053413 &   VB2-Peak \\
                     &                     &  292053413 &  292054635 &   VB2-Dec \\
                     &                     &  292084140 &  292087981 &   {\bf B1}  \\
\hdashline
0604300701 & 2011-03-30 &  417894177 &  417896560 &   {\bf B2}  \\
\hline
\multicolumn{5}{l}{\it This campaign} \\
\hdashline
0743630201 & 2014-08-30 &  525829293 &  525832367 &   {\bf VB3} \\
                     &                     &  525827793 & 525829193 &    VB3-Pre   \\
                     &                     &  525829193 & 525830593 &    VB3-Rise  \\
                     &                     &  525830593 & 525831843 &    VB3-Peak \\
                     &                     &  525831843 & 525832743 &    VB3-Dec   \\
                     &                     &  525832743 & 525834893 &    VB3-Post  \\
                     & 2014-08-31 &  525846661 &  525848532 &   {\bf B3}  \\
\hdashline
0743630301 & 2014-09-01 &  525919377 &  525924133 &   {\bf B4}  \\
\hdashline
0743630501 & 2014-09-29 &  528357937 &  528365793 &   {\bf B5}  \\
\hline
\hline
\end{tabular}
\caption{List of \xmm\ observations and flares considered in this work. 
The flares are divided into two categories, bright (B) and very bright (VB), 
classified according to their total fluence (see Ponti et al. 2015a). 
The different columns show the \xmm\ obsID, flare start and end times in 
Terrestrial Time (TT$\ddag$; see Appendix A) units and flare name, 
respectively. The flare start and end times are barycentric 
corrected (for comparison with multi-wavelength data) and correspond to the 
flare start time (minus 200 s) and flare stop time (plus 200 s) obtained through 
a Bayesian block decomposition (Ponti et al 2015a; please note that 
the time stamps in Ponti et al 2015a are not barycentric corrected). 
Neither a moderate nor weak flare is detected during these \xmm\ observations. 
To investigate the presence of any spectral variability within each 
very bright flare, we extracted three equal duration spectra catching 
the flare rise, peak and decay, with the exception of VB3. In the latter case, 
we optimised the duration of these time intervals according 
to the presence of simultaneous NIR observations (see Fig. \ref{LCp}). } 
\label{obsid}
\end{center}
\end{table} 

We extracted the source photons from a circular region with 
$10~^{\prime\prime}$ radius, 
corresponding to $\sim5.1\times10^4$~AU, or $\sim6.5\times10^5$~$R_S$ 
(Goldwurm et al. 2003; B\'elanger et al.\ 2005; Porquet et al.\ 2008; 
Trap et al.\ 2011; Mossoux et al.\ 2015). 
For each flare we extracted source photons during the time window 
defined by the Bayesian block routine (applied on the EPIC-pn light curve, 
such as in Ponti et al. 2015a), adding 200~s before and after the flare (see 
Tab. \ref{obsid}). 
Background photons have been extracted from the same source regions by 
selecting only quiescent periods. The latter are defined as moments 
during which no flare of \sgras\ is detected by the Bayesian block procedure 
(Ponti et al. 2015a) and additionally leaving a 2~ks gap before the start and 
after the end of each flare. 

Given that all of the observations were taken in Full frame mode, pile-up 
is expected to be an issue only when the count rate exceeds 
$\sim2$ cts/s\footnote{\xmm\ 
User Handbook Issue 2.12, Longinotti et al.\ 2014}. This threshold is above 
the peak count rate registered even during the brightest flares of \sgras. 
This provides \xmm\ with the key advantage of being able to collect pile-up 
free, and therefore unbiased, spectral information even for the brightest flares.

For each spectrum, the response matrix and effective area have been 
computed with the XMM-SAS tasks {\sc rmfgen} and {\sc arfgen}. 
See Appendix A for further details on the \xmm\ data reduction. 

\subsection{\nustar}
\label{sub:nustar}

To study the flare characteristics in the broad X-ray band, 
we analyzed the two {\it NuSTAR} observations (obsID: 30002002002, 
30002002004) taken in fall 2014 in coordination with {\it XMM-Newton}. 
We processed the data using the \nustar\ {\it Data Analysis Software 
NuSTARDAS} v.1.3.1. and HEASOFT v. 6.13, filtered for periods of high 
instrumental background due to SAA passages and known bad 
detector pixels. 
Photon arrival times were corrected for on-board clock drift and 
precessed to the Solar System barycenter using the JPL-DE200 ephemeris.
For each observation, we registered the images with the brightest point 
sources available in individual observations, improving the astrometry 
to $\sim4^{\prime\prime}$.
We made use of the date obtained by both focal plane modules 
FPMA and FPMB.

Four {\it XMM-Newton} flares were captured in the coordinated 
{\it NuSTAR} observations: VB3, B3, B4 and B5.
We extract the {\it NuSTAR} flare spectra using the same flaring intervals 
as determined from the {\it XMM-Newton} data (see Table 1). 
The flare times are barycentric corrected for comparison between 
different instruments. Due to interruption caused by earth occultation, 
{\it NuSTAR} good time intervals (GTIs) detected only a portion of the flares.  
For flare VB3, {\it NuSTAR} captured the first $\sim1215~s$ of the 
full flare, corresponding to pre-, rising- and part of the peak-flare, 
while the dec- and post-flare intervals were missed.
Similarly, part of the rising-flare stage of flare B3 and the middle half 
of flare B4 were captured in the {\it NuSTAR} GTIs. 
Flare B5 was not significantly detected with {\it NuSTAR}, resulting 
in $\sim2\sigma$ detection in the {\it NuSTAR} energy band.

To derive the flare spectra, we used a source extraction region with 
$30^{\prime\prime}$ radius centered on the position of {Sgr~A$^{\star}$}.
While the source spectra were extracted from the flaring intervals, 
the background spectra were extracted from the same region in 
the off-flare intervals within the same observation. 
The spectra obtained by FPMA and FPMB are combined and then 
grouped with a minimum of $3\sigma$ signal-to-noise significance 
per data bin, except the last bin at the high-energy end for which 
we require a minimum of $2\sigma$ significance. 

\begin{table*}
\begin{tabular}{lcccc}                                                                                                      
\hline
\multicolumn{5}{c}{Coordinated {\it NuSTAR} and {\it XMM-Newton} observations and flares detected} \\
{\it NuSTAR}   & $t_{start}$   & Exposure   & Joint {\it XMM-Newton}    & NAME \\
obsID             &                     & (ks)            & obsID                                &           \\
\hline
30002002002     &   2014-08-30 19:45:07  	  &59.79~ks    	  & 0743630201 	  &  VB3, B3 \\
                               &                                                     &                         & 0743630301          &   B4   \\
30002002004  	   &    2014-09-27 17:31:07           &67.24~ks    	  & 0743630501          &   B5   \\
\hline
\end{tabular}
\caption{The different columns show the {\it NuSTAR} obsID, observation 
start time, total exposure, coordinated {\it XMM-Newton} obsID and 
the flares detected in the observation.}  
\end{table*}        					     			     

\subsection{\sinfoni}      
\label{sub:sinfoni}
\subsubsection{Observations and data reduction}

We observed \sgras\ with \sinfoni\ (Eisenhauer et al. 2003; Bonnet et al. 2004) 
at VLT between 30-08-2014 23:19:38 UTC and 31-08-2014 01:31:14 UTC. 
\sinfoni\ is an adaptive optics (AO) assisted integral-field spectrometer mounted 
at the Cassegrain focus of Unit Telescope 4 (Yepun) of the ESO Very Large 
Telescope. 
The field of view used for this observation was $0.8^{\prime\prime} \times 
0.8^{\prime\prime}$, which is divided 
into $64\times32$ spatial pixels by the reconstruction of the pseudoslit into a 3D 
image cube. We observed in H$+$K bands with a spectral resolution of $\sim1800$.  

We accumulated seven spectra (see observation log in Tab. \ref{IRtime}) of 600~s 
each using an object-sky-object observing pattern. 
There are gaps between observations for the 600~s sky exposures, as well as 
a longer gap due to a brief telescope failure during what would have been 
an additional object frame at the peak of the X-ray flare.
In total, we accumulated four sky frames on the sky field (712$^{\prime\prime}$ 
west, $406^{\prime\prime}$ north of \sgras). 
During our observations, the seeing was $\sim$ $0.7^{\prime\prime}$ and 
the optical coherence 
time was $\sim$ 2.5 ms. The AO loop was closed on the closest optical guide star 
($m_R$ = 14.65; $10.8^{\prime\prime}$ east, $18.8^{\prime\prime}$ north 
of \sgras), yielding a spatial resolution 
of $\sim 90 $ mas FWHM at 2.2 $\mu $m, which is $\sim$1.5 times 
the diffraction limit of UT4 in K band.

The reduction of the \sinfoni\ data followed the standard steps. The object 
frames were sky subtracted using the nearest-in-time sky frame to correct 
for instrumental and atmospheric backgrounds. We applied bad pixel correction, 
flat-fielding, and distortion correction to remove the intrinsic distortion 
in the spectrograph. We performed an initial wavelength calibration with calibration 
source arc lamps, and then fine-tuned the wavelength calibration using 
the atmospheric OH lines of the raw frames. Finally, we assembled the data 
into cubes with a spatial grid of 12.5 mas per pixel.

\subsubsection{\sgras\ spectrum extraction}
\begin{figure*}
\includegraphics[height=0.45\textwidth,angle=0]{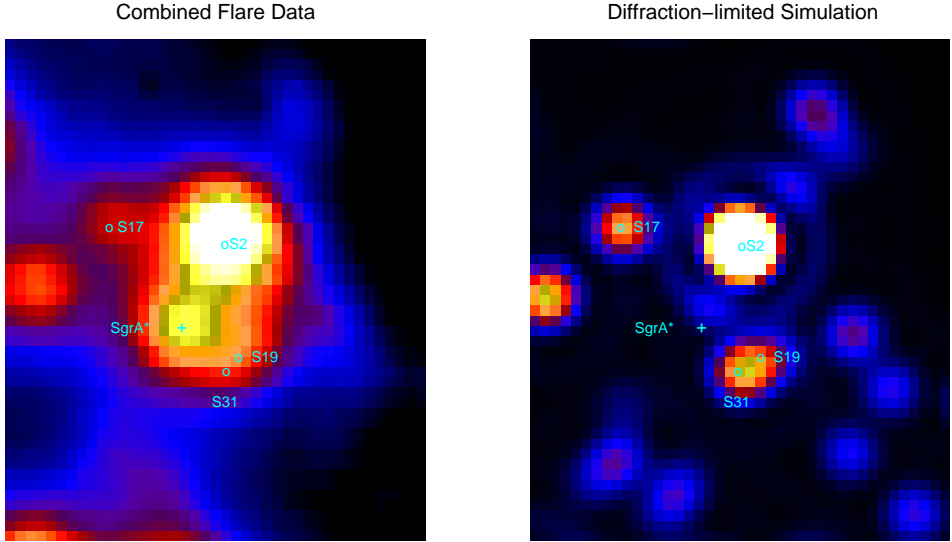}
\caption{The GC as seen with \sinfoni\ (each image is 0.51" 
$\times$ 0.61").
The image on the left is the collapsed image from the spectral range 
2.25-2.35 $\mu$m for the seven data cubes combined. 
The image on the right is a diffraction-limited simulated image of 
the locations of the S-stars in the GC. In both images 
the location of \sgras\ is indicated with a cross, and the four stars 
included in our Gaussian fits for spectrum extraction are labeled 
with circles. The flare is clearly visible above the background 
in the left image.}
\label{ImaNIR}
\end{figure*}

The source spectrum extraction uses a procedure to extract a noisy spectrum 
from \sgras. This can then be binned and the scatter within a bin used 
as an estimate of the error on the flux in that bin.
In each of the seven data cubes, we use a rectangular region of the spatial 
dimensions of size ($0.31^{\prime\prime}$ $\times$ $0.36^{\prime\prime}$) 
centered roughly between \sgras\ and the bright star S2 (0.05" south of S2).
Within this region are four known S-stars, S2, S17, S19, and S31. 
Figure \ref{ImaNIR} shows a combined data cube assembled from 
the seven observations, 
as well as a simulated image of the Galactic centre S-stars. 
The four known stars used in the fitting procedure are labelled in both images. 
We extract $\sim$100 noisy images from the data cube by collapsing 
the cube along the spectral direction (median in the spectral direction) 
in bins with 3.5 nm width (seven spectral channels per bin) in the spectral 
range 2.03-2.39 $\mu$m. 
This initial binning is necessary, as the signal to noise of a single spectral 
channel is not high enough on its own to perform the next step.

In each noisy image, we determine the flux of \sgras\ from a fit with 6 Gaussians 
to the image. Five Gaussians with a common (variable) width describe 
the five sources in each image. 
The sixth Gaussian has a width of 3.5 times wider than the sources, and 
describes the AO seeing halo of the brightest star, S2, which has a K magnitude 
of $\sim$14. 
The seeing halos of the dimmer stars (K magnitude $<$15) are neglected in the fit.
The positions of the four stars relative to one another and to \sgras\ are fixed 
based on the known positions of the stars.
The flux ratios of the four stars are fixed based on previous photometric 
measurements of the stars. 
Note that fixing the flux ratios assumes that the spectral indices of the 
various S-stars are not significantly different, an excellent assumption given 
the strong extinction toward the Galactic center (GC).

The final fit has five free parameters: The overall amplitude of the S-stars, 
the background, the Gaussian width of the sources, the flux ratio of the seeing 
halo/S2, and the flux ratio of \sgras/S2.
This fitting procedure allows a measurement of the variability of \sgras\ in 
the presence of variations (in time and wavelength) in the background, Strehl 
ratio, and seeing.
The result of this procedure is a flux ratio of \sgras/S2 in each of the $\sim$ 100 spectral bins.

We obtain a noisy, color-corrected spectrum of \sgras\ by multiplying a calculated 
spectrum of S2 by the flux ratio \sgras/S2 obtained from the fit in each 
extracted image.
The calculated S2 spectrum used is $\nu S_{\nu}$ for a blackbody with a
temperature of 25,000K, and a stellar radius of 9.3 $R_\odot$, the best fit 
temperature and radius for S2 found in Martins et al. (2009). 
The source is placed at 8.2 kpc (Genzel et al. 2010a) from the Earth. 
This spectrum is normalized to a value of 20 mJy at 2.2 $\mu$m wavelength 
to match previous photometric measurements of S2.
This procedure corrects for the effects of interstellar extinction.
Note that by normalizing the spectrum of S2 to a value of 20~mJy at 2.2~$\mu$m, 
we do not take into account the error on the previous measurements 
of the flux of S2. Since errors on this value result only in an overall error 
on the amplitude and not in the spectral shape, this additional uncertainty 
in the normalization of the spectra is taken into account in the later model fits 
by allowing the overall amplitude of the NIR spectrum to vary and determining 
the effect of this variation on the fit parameters.

To obtain the final NIR data points used for the model fitting in this paper, 
the noisy spectrum is binned into 10 spectral bins (median of the values 
in each bin) of width 35 nm. The error on each point is the standard deviation 
of the sample, or $\sigma/\sqrt{N}$. 
We have tested varying the number of initial spectral samples used to create 
the extracted images used for fitting \sgras/S2, and find that it has almost 
no effect on the final data values and only a small effect on the derived error bars.
We have also tried fixing parameters in the fits to determine their effects 
on the final spectra. 
We tried fixing the FWHM of the Gaussians and the background level (which 
both naturally vary with wavelength) to their median values, and found that 
this affects the spectral index of the final data points by at most a few percents. 

\begin{table*} 
\begin{center}
\small
\begin{tabular}{ c c c c c c }
\hline
Spec & t$_{\rm start}$       & t$_{\rm stop}$     &$\Gamma$     & $F_{2.2\mu m}\dag$      \\
         & BJD$_{\rm TBD}$ & BJD$_{\rm TDB}$&                       & (mJy)                   \\
\hline
IR1 & 2456900.47470 & 2456900.48164 & $1.48\pm0.23$ & $8.87\pm0.10$ \\
IR2 & 2456900.48971 & 2456900.49665 & $1.37\pm0.19$ & $7.91\pm0.07$ \\
IR3 & 2456900.49694 & 2456900.50388 & $1.80\pm0.13$ & $10.09\pm0.07$ \\
IR4 & 2456900.52160 & 2456900.52855 & $1.76\pm0.21$ & $7.52\pm0.06$ \\
IR5 & 2456900.53675 & 2456900.54370 & $3.71\pm0.48$ & $3.12\pm0.07$ \\
IR6 & 2456900.54398 & 2456900.55093 & $2.72\pm0.23$ & $4.23\pm0.04$ \\
IR7 & 2456900.55914 & 2456900.56608 & $2.56\pm0.21$ & $7.07\pm0.06$ \\
\hline
\end{tabular}
\caption{The second and third column show the Barycentre corrected start 
and end times of each of the seven \sinfoni\ spectra. Times are Barycentric 
Julian Dates in Barycentric Dynamical Time. The last two columns show the 
test fit photon index ($\Gamma$) and flux once the \sinfoni\ data are fitted 
with a simple power-law model. \dag The error bars are statistical only and 
they do not include systematic effects due to spectral extraction. The effect 
of systematics is treated in \S 5.3.4.}
\label{IRtime}
\end{center}
\end{table*}

\section{X-ray obscuration and mean X-ray properties of bright flares}
\label{brightxmm}

We started the study of \sgras's emission by investigating the properties 
of the absorption-scattering layers that distort its spectrum. 

\subsection{Dust scattering} 
\label{IntroDSH}

Scattering on dust grains along the line of sight can have a significant impact 
on the observed X-ray spectra (Predehl \& Schmitt 1995; Smith et al. 2016). 
The main effect of dust scattering is to create a halo around the source, 
by removing flux from the line of sight. Both the flux in the halo and its size 
decrease with energy (with a dependence of $\propto E^{-2}$ and 
$\propto E^{-1}$, respectively) as a consequence of the probability 
of scattering that drops steeply with energy. If the events used to 
extract the source photons are selected from a small region containing only 
a small part of the halo, such as typically the case for X-ray observations 
of \sgras, then the spectral shape will be distorted by the effects of 
dust scattering. 
Whenever the distortions are not accurately accounted for, this will 
cause significant biases in the measured absorption column densities, 
source brightness and spectral slopes (see appendix \ref{SecDSH}). 

Frequently used models, aimed at mitigating the effects of dust 
scattering, are: {\sc dust} 
(Predehl \& Schmitt 1995); {\sc scatter} (Porquet et al. 2003; 2008; 
Dodds-Eden et al. 2009) and; {\sc dustscat} (Baganoff et al. 2003; 
Nowak et al. 2012). In all these models the dust optical depth and therefore 
the magnitude of the correction, is assumed to be proportional to the 
X-ray absorbing column density, with a factor derived from Predehl 
\& Schmitt (1995). The underlying assumption is that the dust properties 
(e.g., dust to gas ratio, size distribution, composition, etc.) towards 
\sgras\ are equal to the average estimate derived from the study 
of all the Galactic sources considered in the work of Predehl \& Schmitt (1995).
After considering the limitations of this approach, we decided to use 
a completely different method. 

Thanks to the analysis of all the \xmm\ and \chandra\ observations 
of the GC, Jin et al. (2016) just completed the accurate characterisation 
of the dust scattering halo towards \axj, a bright X-ray binary located 
only $\sim1.45^{\prime}$ from \sgras. The authors deduced that 
$74\pm7$~\% of the dust towards \axj\ resides in front of the GC 
(e.g. in the spiral arms of the Galaxy).
Moreover, the detailed modelling of the dust scattering halo allowed 
Jin et al. (2016) to provide an improved model of the spectral distortions 
generated by the dust scattering ({\sc fgcdust}), without the 
requirement to assume fudge scaling factors. We therefore decided 
to fit \sgras's spectrum with the {\sc fgcdust} model, implicitly 
assuming that the dust has similar properties along the line of sight 
towards \sgras\ and the foreground component in the direction of \axj. 
This is corroborated by the study of the radial and azimuthal 
dependence of the halo. In fact, the smoothness of the profile indicates 
that the foreground absorption has no major column density variations 
within $\sim100-150^{\prime\prime}$ from \axj\ (Jin et al. 2016). 
Further details on the spectral distortions (and their correction) introduced 
by dust scattering are discussed in Appendix \ref{SecDSH}. 

\subsection{Foreground absorption towards the bright sources within
the central arcmin}
\label{ISMsources}

We review here the measurements of the X-ray column density 
of neutral/low-ionised material along the line of sights towards 
compact sources located close to \sgras. 
Due to the variety of assumptions performed in the different works 
(e.g., absorption models, abundances, cross sections, dust scattering 
modelling, etc.), we decided to refit the spectra to make all measurements 
comparable with the abundances, cross sections and absorption models 
assumed in this work (see \S 2.1; Fig. 5 and Tab. 2 of Ponti et al. 2015 
and Ponti et al. 2016). \\
{\bf \axj\ located at $\sim1.45^{\prime}$ from \sgras}\\ 
\axj\ is a dipping and eclipsing neutron star low mass X-ray binary (Ponti 
et al. 2016). Such as typically observed in high inclination low mass 
X-ray binaries, \axj\ shows both variable ionised and neutral local 
absorption (Ponti et al. 2015). 
The total neutral absorption column density towards \axj\ has been 
measured by Ponti et al. (2015b). We re-fitted those spectra of \axj\ with 
the improved correction for the dust scattering distortions. By considering 
that only $74\pm7$~\% of the dust towards \axj\ resides in front of 
the GC (Jin et al. 2016), we measured a total column density in the 
foreground component\footnote{The large 
uncertainty in this measurement is driven by the uncertainty in 
the determination of the fraction of column density in the foreground 
component. } of $N_H=(1.7\pm0.2)\times10^{23}$~cm$^{-2}$. 
We note that the halo associated to the foreground component 
is still detected at radii larger than $r>100^{\prime\prime}$ 
(Jin et al.\ 2016), therefore from a radius more than ten times 
larger than the one chosen to extract \sgras's photons, 
indicating that a careful treatment of the distortions introduced 
by dust scattering is essential. 
\\
{\bf SWIFT~J174540.7-290015 located at $\sim16^{\prime\prime}$ 
from \sgras}\\ 
A deep \xmm\ observation performed during the recent outburst of \ssv\ 
(Reynolds et al. 2016), allowed Ponti et al. (2016) to measure the 
column density along this line of sight 
and to find $N_H=(1.70\pm0.03)\times10^{23}$ cm$^{-2}$, by fitting the spectrum
with the sum of a black body plus a Comptonisation component\footnote{The 
authors find marginal evidence for sub-Solar iron abundance, 
suggesting that iron is depleted into dust grains. The detailed investigation 
of the metal abundances is beyond the scope of this paper. }. 
By applying the improved modelling of the dust scattering halo to the same 
data, we measured a column density of $N_H=(1.60\pm0.03)\times10^{23}$ 
cm$^{-2}$.\\
{\bf SGR~J1745-2900 located at $\sim2.4^{\prime\prime}$ from \sgras}\\
\sgr\ is a magnetar located at a small projected distance from \sgras\ 
(Mori et al. 2013; Kennea et al. 2013), 
and it is most likely in orbit around the supermassive BH (Rea et al. 2013).
Coti-Zelati et al. (2015) fitted the full \xmm\ and \chandra\ dataset available 
on \sgr, without considering the effects of the dust scattering halo, and 
found $N_H=(1.90\pm0.02)\times10^{23}$ cm$^{-2}$ for \chandra\ and 
$N_H=(1.86^{+0.05}_{-0.03})\times10^{23}$ cm$^{-2}$ for \xmm. 
We refitted the \xmm\ dataset at the peak of emission (obsid: 0724210201),
using as background the same location when the magnetar was 
in quiescence and considering the improved dust model, and 
we obtained $N_H=(1.69^{+0.17}_{-0.10})\times10^{23}$ cm$^{-2}$. 

Radio observations of the pulsed emission from \sgr\ allowed Bower et al. 
(2014) to provide a full characterisation of the scattering properties of 
the absorption. The authors found the obscuring-scattering layer 
to be located in the spiral arms of the Milky Way, most likely at 
a distance $\Delta=5.8\pm0.3$~kpc from the GC (however a uniform scattering 
medium was also possible). Moreover, the source sizes at different 
frequencies are indistinguishable 
from those of \sgras, demonstrating that \sgr\ is located behind 
the same scattering medium of \sgras. 

We note that the column densities of absorbing material along 
the line of sights towards SWIFT~J174540.7-290015, \sgr\ and 
the foreground component towards \axj\ are consistent (to an 
uncertainty of $\sim2-10$~\%) within each other. Of course, 
the neutral absorption towards these accreting sources might even, 
in theory, be local and variable (e.g. Diaz-Trigo et al. 2006; 
Ponti et al. 2012; 2016b), however the similar values observed 
in nearby sources indicate a dominant ISM origin. We note 
that the location of the scattering medium towards \sgras\ and 
the foreground component of \axj\ are also cospatial. 
This suggests that all these sources are absorbed by a common, 
rather uniform, absorbing layer located in the spiral arms of 
the Milky Way (Bower et al. 2014; Jin et al. 2016). This result is also 
in line with the small spread, of the order of $\sim10$~\%, in the extinction 
observed in NIR towards the central $\sim20^{\prime\prime}$ of the 
Galaxy (Sch\"odel et al. 2010; Fritz et al. 2011). 
Indeed, for a constant dust to gas ratio, this would induce a spread in 
the X-ray determined $N_H$ of a similarly small amplitude. 
In addition to this layer, \axj\ also shows another absorbing 
component, located closer to the source, possibly associated 
either with the clouds of the central molecular clouds or a 
local absorption (Jin et al. 2016). 

Studies of the scattering sizes from large scale ($\sim2^\circ$) 
low frequency radio maps also agree with the idea that 
the intervening scattering in the GC direction is composed 
of two main absorption components, one uniform on a large 
scale and one patchy at an angular scale of $\sim10^{\prime}$ 
and with a distribution following the clouds of the central 
molecular zone (Roy 2013). 

\subsection{The mean spectra of the \xmm\ very bright flares}

We extracted an EPIC-pn and -MOS spectra for each of the bright and 
very bright flares detected by \xmm. We used a Bayesian block decomposition 
of the EPIC-pn light curve to define the start and end flare times (see Tab. 
\ref{obsid}). 
We fitted each spectrum with a power-law model modified by neutral 
absorption (see \S \ref{Assumptions}) and by the contribution from the 
dust scattering halo ({\sc fgcdust * TBnew * power-law} in {\sc XSpec}). 

Each spectrum is well fitted by this simple model (see Tab. \ref{bestfitFlares}). 
In particular, the column density of absorption material and the photon 
index are consistent with being the same between the different flares 
and consistent with the values observed in nearby sources (see \S 
\ref{ISMsources}). This agrees with the idea that most of the neutral absorption 
column density observed towards \sgras\ is due to the interstellar medium 
(ISM). If so, the absorption should not vary significantly over time (see 
Tab. \ref{bestfitFlares}). 
Therefore, we repeat the fit of the spectra assuming that the three very 
bright flares are absorbed by the same column density of neutral material. 
The three spectra are well described by this simple model 
(see Tab. \ref{bestfitFlares}), significantly reducing the uncertainties. 
The best fit spectral index is $\Gamma_{VB123}=2.20\pm0.15$, while the 
column density is: $N_H=(1.59\pm0.15)\times10^{23}$~cm$^{-2}$. 

This value is fully consistent with the one observed towards the foreground 
component towards bright nearby X-ray sources, reinforcing the suggestion 
that the column density is mainly due to the ISM absorption. 
For this reason hereafter we will fix it to the most precisely constrained 
value $N_H=1.60\pm0.03\times10^{23}$~cm$^{-2}$ (Ponti et al. 2016b). 
The resulting best fit photon index with this value of $N_H$ is 
$\Gamma_{VB123_{FixN_{H}}}=2.21\pm0.09$. 
\begin{table} 
\begin{center}
\scriptsize
\begin{tabular}{ c c c c c c }
\hline
\multicolumn{5}{c}{\bf Absorbed power-law fit to X-ray spectra} \\
\hline
Name & $N_H$     & $\Gamma_X$    & Flux$_{2-10}$           & $\chi^2$/dof \\
\hline
VB1   & $1.6\pm0.2$ & $2.2\pm0.3$ & $9.6^{+7}_{-4}$       & 89.9/114 \\
VB2   & $1.6\pm0.3$ & $2.3\pm0.4$ & $5.0^{+5}_{-2.4}$    & 89.3/98 \\
VB3   & $1.6\pm0.3$ & $2.3\pm0.3$ & $7.6^{+7.1}_{-3.4}$ &127.2/117 \\
VB123\dag & $1.59\pm0.15$ & $2.20\pm0.15$ &                & 302.8/331 \\  
VB123$_{FixN_{H}}$\ddag & $1.6$ & $2.21\pm0.09$ && 302.8/332 \\  
VB3$_{XMM+Nu}$\ddag     & $1.6$ & $2.27\pm0.12$ &$7.5\pm1.5$&141.4/133 \\
\hline
\end{tabular}
\caption{Best fit parameters of the fit of the very bright flares of \sgras. 
Column densities are given in units of $10^{23}$~cm$^{-2}$ and 
the absorbed fluxes are in units of $10^{-12}$ erg cm$^{-2}$ s$^{-1}$. 
\dag The VB123 flare indicates the average of VB1+VB2+VB3. 
\ddag The VB123$_{FixN_{H}}$ shows the best fit results of flare 
VB123, once the column density of neutral absorbing material 
has been fixed. The VB3$_{XMM+Nu}$ shows the best fit results of 
flare VB3 (by fitting both \xmm\ and \nustar\ data), once the column 
density has been fixed. }
\label{bestfitFlares}
\end{center}
\end{table} 

\section{Mean properties of VB3}
\label{multiwave}

We investigate here the mean properties of a very bright flare (VB3, see Ponti 
et al. 2015a) during which, for the first time, simultaneous time-resolved 
spectroscopy in NIR and X-rays has been measured. 

\subsection{X-ray (\xmm\ and \nustar) mean spectra of VB3}

\begin{figure}
\hspace{-0.5cm}
\includegraphics[height=0.47\textwidth,angle=-90]{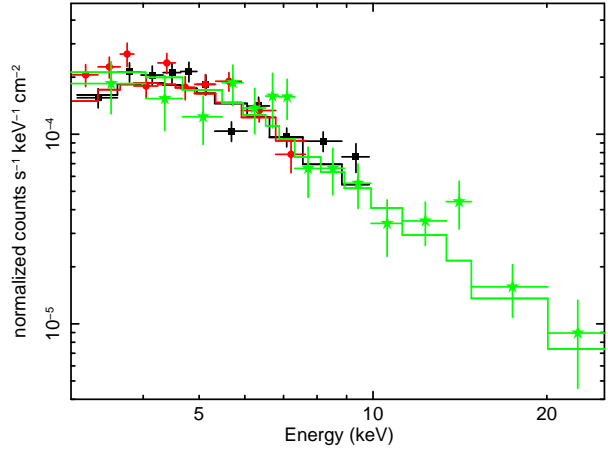}
\caption{Mean X-ray spectrum of VB3. The black squares, red circles and green 
stars show the EPIC-pn, combined EPIC-MOS and combined \nustar\ 
spectra, respectively. The combined \xmm\ and \nustar\ spectra greatly 
improve the determination of the X-ray slope. The data are fitted with 
an absorbed power law model, which takes into account the distortions 
induced by the dust scattering (see text for more details). }
\label{MSpec}
\end{figure}

We first simultaneously fitted the \xmm\ (pn and both MOS) and \nustar\ 
mean spectra of VB3 (see Fig. \ref{MSpec}, Tab. \ref{obsid}) with 
an absorbed power law model. 
The \nustar\ data cover only part of the flare, missing the decaying flank 
of the flare, therefore probing different stages of a variable phenomenon. 
We accounted for this by allowing the fit to have different power-law 
normalisation between the \nustar\ and \xmm\ spectra\footnote{The 
normalisations are, however, consistent between the two instruments.}. 
The best fit photon index is: $\Gamma_{XMM+Nu}=(2.27\pm0.12)$ 
and absorbed 2-10 keV flux $F_{2-10}=7.5\pm1.5\times10^{-12}$ erg 
cm$^{-2}$ s$^{-1}$. The spectra were very well fit by this simple model 
with $\chi^2=141.4$ for 133 dof.  

We investigated for the presence of a possible high-energy cut-off 
(or high-energy spectral break) by fitting the spectra with an absorbed 
broken power-law model. 
We fixed the photon index of the lower energy power-law slope 
to $\Gamma_{VB123_{FixN_{H}}}=2.20$ (the best fit value of the 
simultaneous fit of VB1+VB2+VB3, see Tab. \ref{bestfitFlares}).
No significant improvement was observed ($\chi^2=141.2$ for 132 dof). 

\subsection{Multi-wavelength mean spectra of VB3}
\label{evolution}

We then extended our investigation by adding the NIR spectra. 
Multiple \sinfoni\ spectra (e.g., IR2, IR3 and IR4) have been accumulated
during the duration of the X-ray emission of the VB3 
flare\footnote{Defined on the basis of the X-ray light curve, 
therefore it represents the full duration of the X-ray flare. } (see Fig. \ref{LCp}). 
Therefore, we created the mean spectrum from these NIR spectra 
and fitted these simultaneously with the mean X-ray spectra of VB3 
(see Tab. \ref{IRtime} and Fig. \ref{LCp}). 

\subsubsection{Single power-law (plain Synchrotron)} 
\label{PureSync}

We started fitting the mean spectrum from NIR to hard X-ray with a simple 
power-law model, as expected in the case of plain Synchrotron emission 
(Dodds-Eden et al. 2009). 
The best fit photon index was $\Gamma=2.001\pm0.005$ (see Tab. 
\ref{meanSp} and Fig. \ref{PLCool}). 
However, this very simple model provided us with an unsatisfactory 
result ($\chi^2=189.7$ for 142 dof; Tab. \ref{meanSp}). This is mainly 
driven by the different slopes observed in the NIR and X-ray bands 
(see Fig. \ref{SED} and \ref{PLCool}). 
We therefore concluded that a plain Synchrotron model is ruled out. 

\subsubsection{Broken power-law model (BPL, phenomenological model)} 
\label{BPL}

We then performed a phenomenological description of the data 
with a broken power-law (BPL) model. We observed a significant improvement 
and an acceptable description of the spectrum by fitting the data with 
this model, where the NIR and X-ray slopes were 
free to vary ($\chi^2=154.9$ for 142 dof, $\Delta\chi^2=34.8$ for the 
addition of 2 dof, corresponding to an F-test probability of 
$\sim7\times10^{-7}$; Tab. \ref{meanSp}, Fig. \ref{PLCool}). 
The resulting best fit NIR and X-ray photon indexes are 
$\Gamma_{NIR}=1.7\pm0.1$ and $\Gamma_X=2.27\pm0.12$, 
respectively (Tab. \ref{meanSp}). 
The spectral steepening $\Delta\Gamma=0.57\pm0.15$ 
($\pm0.09$ at 1-$\sigma$) 
is slightly steeper, but fully consistent with the value expected in the 
Synchrotron scenario in the presence of a cooling break 
($\Delta\Gamma=0.5$), strongly suggesting this latter scenario as 
the correct radiative mechanism. 
\begin{table} 
\begin{center}
\footnotesize
\begin{tabular}{ c c c c c c c }
\hline
\multicolumn{6}{c}{\bf VB3 mean spectrum} \\
\hline
                         & Single PL                & BPL                             & TSSC                                    & PLCool                  \\
\hline
$\Gamma_{NIR}$& $2.001\pm0.005$  & $1.7\pm0.1$                &                                               & $1.74\pm0.08$     \\
$\Gamma_X$    &                               & $2.27\pm0.12$            &                                               &                              \\
$\Delta\Gamma$&                              & $0.57\pm0.15$            &                                               & 0.5                        \\
Log(B)               &                               & $0.94\pm0.16$            & $4.0\pm0.4$                         & $0.94\pm0.16$      \\
$\Theta_e$        &                               &                                     & $9\pm4$                               &                               \\
Log($N_e$)       &                               &                                     & $39.5\pm0.5$                       &                               \\
Log(R$_F$)       &                               &                                     & $-3.5\pm0.5$                       &                                \\
\hline
$\chi^2$/dof        & 189.7/142             & 154.9/140                    & 162.7/139                              & 156.8/141                \\
\hline
\end{tabular}
\caption{Best fit parameters of the mean spectrum of VB3 
with the Single PL (plain synchrotron), BPL (broken power-law), 
TSSC (thermal synchrotron self Compton) and PLCool (power-law 
cool) models. 
See \S \ref{multiwave} for a description of the parameters. }
\label{meanSp}
\end{center} 
\end{table} 

\subsubsection{Thermal Synchrotron Self Compton (TSSC)}
\label{TSSC}

Before fitting the VB3 mean spectrum with a Synchrotron model with 
a cooling break, we considered an alternative interpretation, where 
the NIR band is produced via synchrotron radiation by a thermal 
distribution of electrons. Moreover, the same population of electrons 
generates via inverse Compton the high-energy (e.g. X-ray) emission 
(see e.g. Dodds-Eden et al. 2011). We called this model thermal 
Synchrotron self Compton (TSSC). 
The free parameters in this model are: $B$, the strength of the magnetic 
field; $\theta_E$, the dimensionless electron temperature (defined 
as $\theta_E=\frac{kT_e}{m_e c^2}$, where $k$ is the Boltzman constant, 
$T_e$ is the temperature of the thermal electrons, $m_e$ is the 
electron mass and $c$ is the speed of light); $N$, the total 
number of NIR synchrotron emitting electrons; and $R_F$, the size of 
the region containing the flaring electrons, controlling the photon density 
of the seed photons. The very short variability time-scale (of the order 
of $10^2$~s) suggests a very compact source with a size of the order 
of (or smaller than) a few Schwarzschild radii, likely located within 
or in the proximity of the hot accretion flow of \sgras. 
Radio and sub-mm observations constrain the physical 
parameters of the steady emission from the inner hot accretion flow 
(within the central $\sim10$~R$_S$) to be $B\sim10-50$~G, 
$T_e\sim10^{10}$~K, $\gamma_e\sim10$; $n_e\sim10^6$~cm$^{-3}$ 
(see \S~\ref{intro}; Loeb \& Waxman 2007; Genzel et al. 2010). 
These are likely the pre-flare plasma conditions. 

The TSSC model provides an acceptable fit to the data 
($\chi^2=162.7$ for 139 dof; see Fig. \ref{PLCool})\footnote{The fits 
have been performed in {\sc matlab}, 
implementing the equations reported in Dodds-Eden et al. (2009) 
and references therein. The best fit was computed through a $\chi^2$ 
minimisation technique. The parameter space to determine the 
uncertainties on the best fit parameters has been explored through 
a Bayesian Markov Chain Monte Carlo approach. }. 
However, as observed in previous very bright flares (Dodds-Eden et 
al. 2009), the best fit parameters of this model are very different 
from the reasonable range expected to be present in the accretion 
flow of \sgras. Indeed, this model produces the flare via a magnetic 
field with a staggering intensity of $Log(B)=4.0\pm0.4$~G, about 
three orders of magnitude larger than the magnetic field intensity 
within the steady hot accretion flow of \sgras, on a population 
of ``not-so-energetic" ($\theta_E=9\pm4$) electrons. 
Moreover, in order to make the inverse Compton process 
efficient enough to be competitive to synchrotron, the electron 
density has to be as high as $n_e=10^{13}$~cm$^{-3}$, about 
seven orders of magnitude higher than in the accretion flow. 
This appears unlikely. 
The total number of TSSC emitting electrons is constrained 
by the model to be $Log(N_e)=39.5\pm0.5$, therefore to reach 
such an excessively high electron density, the size of 
the emitting region has to be uncomfortably small, 
$Log(R_F/R_S)=-3.5\pm0.5$. Such a source would be 
characterised by a light crossing time of the order of only 
$\sim10$~ms. Indeed, variability on such time-scales are typically 
observed in accreting X-ray binaries (e.g., Belloni et al. 2002; 
De Marco et al. 2015), where the system is $\sim10^6$ times 
more compact than in \sgras\ (Czerny et al. 2001; Gierlinski et 
al. 2008; Ponti et al. 2012b), 
while \sgras's  power spectral density appears dominated by 
variability at much larger time-scales (Do et al. 2009; Meyer et al. 
2009; Witzel et al. 2012; Hora et al. 2014)\footnote{ 
An excessively compact source with such high densities 
and magnetic field is hardly achievable event through compression 
of a fraction of the quiescent electrons. The quiescent density of 
electrons ($n_e=10^6$~cm$^{-3}$) dictates that $N_e=10^{39.5}$ 
electrons are contained within a sphere of $\sim0.07$~R$_S$, 
that therefore would need to be compressed by 2.3 orders of 
magnitude to reach the required TSSC source size and density. 
We note that, assuming conservation of the magnetic flux, 
the magnetic field strength would rise to $Log(B)\sim6.2$, 
two orders of magnitude higher than the best fit value. 
The magnetic energy would also rise, but would still be 2-4 orders 
of magnitude smaller than the one required to power the flare. }. 
Again, this appears as a weakness of this model. 

As already discussed in Dodds-Eden et al. (2009; 2010; 2011) and 
Dibi et al. (2014; 2016), these physical values are different by 
several orders of magnitude from the ones observed in quiescence 
and, therefore they appear unlikely. With this study we show that the same 
"unlikely" physical parameters are not only observed during VB2 
(the flare analysed by Dodds-Eden et al. 2009), but also during the very 
bright flare considered here (VB3). This confirms that this model 
produces unreasonable parameters for part (if not all) of the bright flares. 

\subsubsection{Synchrotron emission with cooling break (PLCool)} 
\label{SecPLCool}

In this scenario, the synchrotron 
emission is produced by a non-thermal distribution of relativistic electrons, 
embedded in a magnetic field with strength $B$, therefore they radiate 
synchrotron emission. At the acceleration site, the injected electrons are 
assumed to have a power-law distribution in $\gamma_e$ with index 
$p$ (i.e. $N(\gamma_e)\propto\gamma_e^{-p}$), 
defined between $\gamma_{min}$ and $\gamma_{max}$ ($\gamma_e$ being 
the electron Lorentz factor). We assume as lower boundary 
$\gamma_{min}=10$, supposing that the radiating electrons are accelerated 
from the thermal pool producing the sub-mm peak in quiescence (Narayan et al. 
1998; Yuan et al. 2003). We also assume that at any point during the flare, 
the engine is capable of accelerating electrons to $\gamma_{max}>10^6$, 
so that they can produce X-ray emission via synchrotron radiation 
(this appears as a less reliable assumption and indeed an alternative to 
this scenario will be discussed in \S~\ref{SecDiscPLCoolEv}). 

A well known property of high-energy electrons radiating via synchrotron 
emission is that they cool rapidly, quickly radiating their energy on 
a time-scale $t_{cool}=220 (B/50~G)^{-3/2} (\nu/10^{14}~Hz)^{-1/2}$~s 
(where $\nu$ is the frequency of the synchrotron emitted radiation; 
see Pacholczyk 1970). In particular, higher energy electrons 
cool faster than the NIR ones. The competition between synchrotron 
cooling and particle escape from the acceleration zone then generates 
a break in the synchrotron spectrum at a frequency: 
$\nu_{br}=2.56(B/30~G)^{-3}(t_{esc}/300~s)^{-2}\times10^{14}$~Hz. 
Furthermore, in case of continuous acceleration, a steady solution exists 
where the slope of the power-law above the break is steeper by 
$\Delta\Gamma=0.5$ (Kardashev et al. 1962) than the lower energy 
power-law\footnote{Synchrotron radiation from the cooling power-law 
distribution of electrons (with electron power-law index $p$) generates 
a spectrum $\nu F_\nu \propto \nu^{(3-p)/2}$ at frequencies lower 
than the cooling break and $\nu F_\nu \propto \nu^{(2-p)/2}$ above. 
This implies that $p$ relates to the photon index $\Gamma$ such as: 
$\Gamma=\frac{p+1}{2}$ below and $\Gamma=\frac{p+2}{2}$ at 
frequencies above the cooling break. }. 
Following the nomenclature of Dodds-Eden et al. (2009), 
we call this model "PLCool". The free parameters of the PLCool model 
are: $B$; $p$; and the normalisation.

\begin{figure}
\includegraphics[height=0.49\textwidth,angle=90]{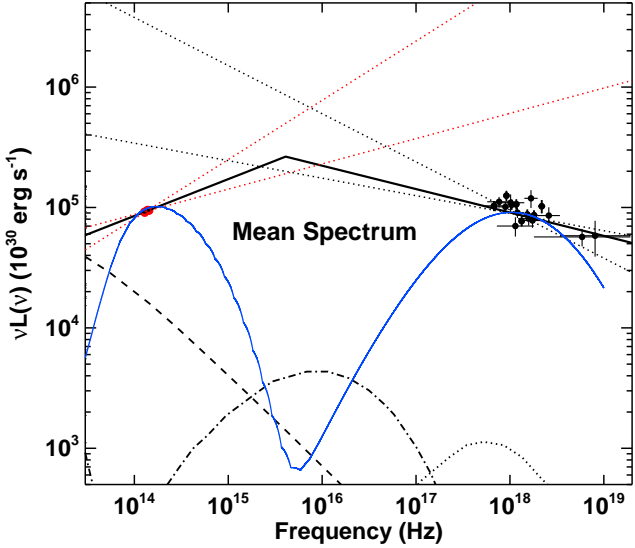}
\caption{The red and black points show the mean NIR (\sinfoni)
and X-ray (\xmm\ and \nustar) emission during the VB3 flare. 
The dotted red and black straight lines show the uncertainties 
on the determination of the NIR and X-ray power-law slope 
(with model BPL), respectively. The solid line shows the best fit 
PLCool model that imposes $\Gamma_{X}=\Gamma_{NIR}+0.5$. 
The X-ray slope is slightly steeper ($\Delta\Gamma=0.57\pm0.09$, 
1$\sigma$), although consistent with the predictions of the PLCool 
model. Both X-ray and NIR data and models have been corrected
for absorption and the effects of dust scattering halo. 
The blue solid line shows the best fit TSSC model (\S 5.3.3).
For a description of the other lines see Fig. \ref{SED}. }
\label{PLCool}
\end{figure}

As described in \S \ref{BPL} a broken power-law model provides 
an excellent fit to the mean VB3 multi-wavelength spectrum 
($\chi^2=154.9$ for 140 dof). 
In particular, we note that the difference between the NIR and X-ray 
photon indices $\Delta\Gamma=0.57\pm0.15$ ($\pm0.09$ at 1-$\sigma$) 
is consistent with the value expected by the PLCool model ($\Delta\Gamma=0.5$; 
due to synchrotron emission with continuous acceleration). 
Indeed, imposing such spectral break ($\Gamma_{X}=\Gamma_{NIR}+0.5$), 
the fit does not change significantly ($\chi^2=156.8$ for 141 dof), with the 
photon index $\Gamma_{NIR}=1.74\pm0.08$ and the break at 
$0.04^{+0.12}_{-0.03}$~keV ($B=8.8^{+5.0}_{-3.0}$~G; Fig. \ref{PLCool}). 
We note that the PLCool model provides a significantly better fit 
($\chi^2=156.8$) than the TSSC model ($\chi^2=161.3$) despite having 
two fewer free parameters (Tab. \ref{meanSp}). 

To investigate the effects of potential uncertainties on the normalisation 
of the NIR emission, we artificially increased (and decreased) 
the \sinfoni\ spectrum by a factor 1.25 (and 0.75). The statistical quality 
of the fit does not change ($\chi^2=156.8$ for 141 dof, in all 
cases), and neither does the best fit photon index, $\Gamma_{NIR}=1.74\pm0.08$. 
As expected, the main effect of the higher (lower) NIR normalisation is 
to shift the break towards lower (higher) energies, i.e. 
$E_{br}=0.018^{+0.071}_{-0.014}$~keV ($E_{br}=0.046^{+0.20}_{-0.037}$~keV), 
corresponding to $B=11.7^{+8.5}_{-3.3}$~G ($B=8.5^{+5.5}_{-3.0}$~G).

\section{Evolution during VB3}
\label{SecSpEvol}

\subsection{Light curves of VB3}

\begin{figure}
\includegraphics[height=0.37\textwidth,angle=0]{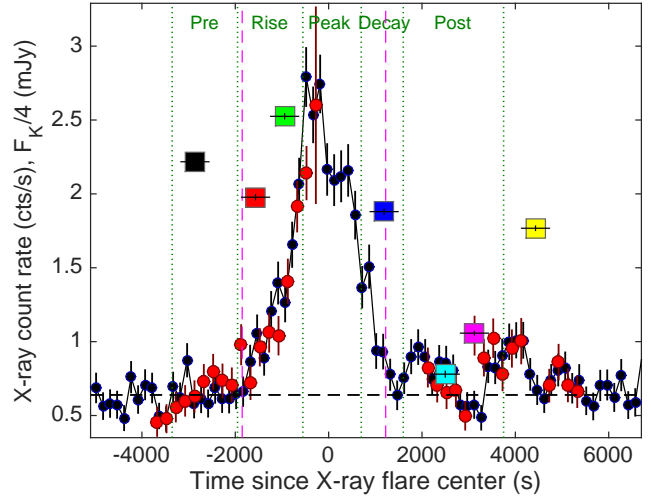}
\caption{The black and red points show the \xmm\ 2-10 keV (sum of the three 
EPIC cameras) and \nustar\ 3-20~keV light curve of \sgras's flare VB3 
(Ponti et al. 2015), respectively. A constant rate of 0.13 cts s$^{-1}$
has been subtracted from the \nustar\ light curve for display purposes,
to take into account the different contribution of the diffuse and quiescent 
emission (gaps in the \nustar\ light curve are due to Earth occultation). 
The squares show the extinction corrected \sinfoni\ light curve 
of \sgras\ during the VB3 flare. Each point corresponds to a NIR 
spectrum integration time of 600~s. The y-axis reports the observed 
renormalised (divided by 4 for display purposes) flux density at 
$2.2~\mu$m in mJy units. The black dashed line indicates the level 
of the ``non-flare" (quiescent in X-rays) emission. The pink dashed lines indicate 
the start and end of the \xmm\ VB3 flare, as indicated by the Bayesian 
block decomposition (see Ponti et al. 2015a). Excess X-ray emission 
is observed $\sim2000$ and $\sim4000$~s after the X-ray flare peak. 
The dotted green lines show the intervals 
for the integration of the pre-, rise, peak, decrease and post-flare 
spectra during VB3. The zero point of the abscissa corresponds 
to $525831144.7$~s (TT$_{\rm TBD}$) and 2456900.50784~day 
(BJD$_{\rm TBD}$), respectively. } 
\label{LCp}
\end{figure}

The black and red points in Fig. \ref{LCp} show \xmm\ and \nustar\ 
light-curves of VB3 in the 2-10~keV and 3-20 keV bands, respectively. 
The black dashed line indicates the level of diffuse and quiescent 
emission observed by \xmm. For display purposes, we subtracted a 
constant rate of 0.13~cts~s$^{-1}$ from the \nustar\ light curve, to take 
into account the different contribution of the diffuse and quiescent 
emission. The squares in Fig. \ref{LCp} show the NIR light-curve as 
observed with \sinfoni. Despite the sparse sampling of the light curve 
allowed by the \sinfoni\ integral field unit, it is clear from Fig. \ref{LCp} 
that for VB3 the NIR flare lasts longer than the X-ray one. 
In particular, the NIR flare is already in progress during our first \sinfoni\ 
integration, $\sim10^3$~s before the start of the X-ray flare and it is 
still in progress at the end of IR4, with a duration longer than $3.4$~ks  
(see Fig. \ref{LCp}, Tab. \ref{obsid} and \ref{IRtime}).
This is not surprising, indeed, a similar trend has already been observed 
by Dodds-Eden et al. (2009) and Trap et al. (2011) in the only other very 
bright flare with simultaneous NIR coverage (see Fig. 3 of Dodds-Eden et al. 2009). 
The pink dashed lines in Fig. \ref{LCp} indicate the start and end 
of the \xmm\ VB3 flare as determined by the Bayesian block 
decomposition (see Ponti et al. 2015a).
The dotted green lines indicate the periods during which VB3-Pre, VB3-Rise, 
VB3-Peak, VB3-Dec and VB3-Post, have been integrated (Tab. \ref{obsid}). 

\subsection{NIR spectral evolution during VB3}
\label{NIR}

\begin{figure}
\includegraphics[height=0.35\textwidth,angle=0]{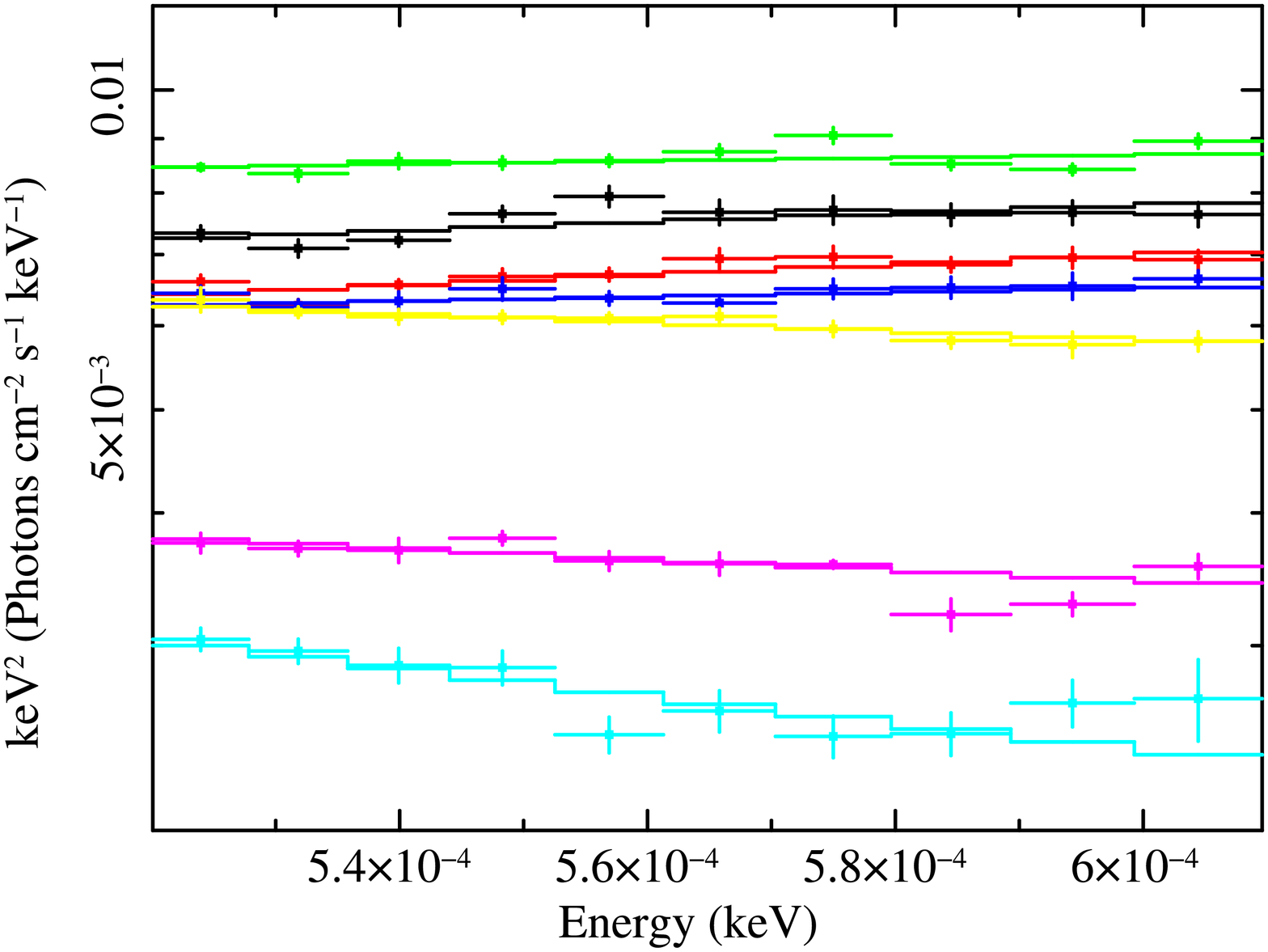}
\includegraphics[height=0.34\textwidth,angle=0]{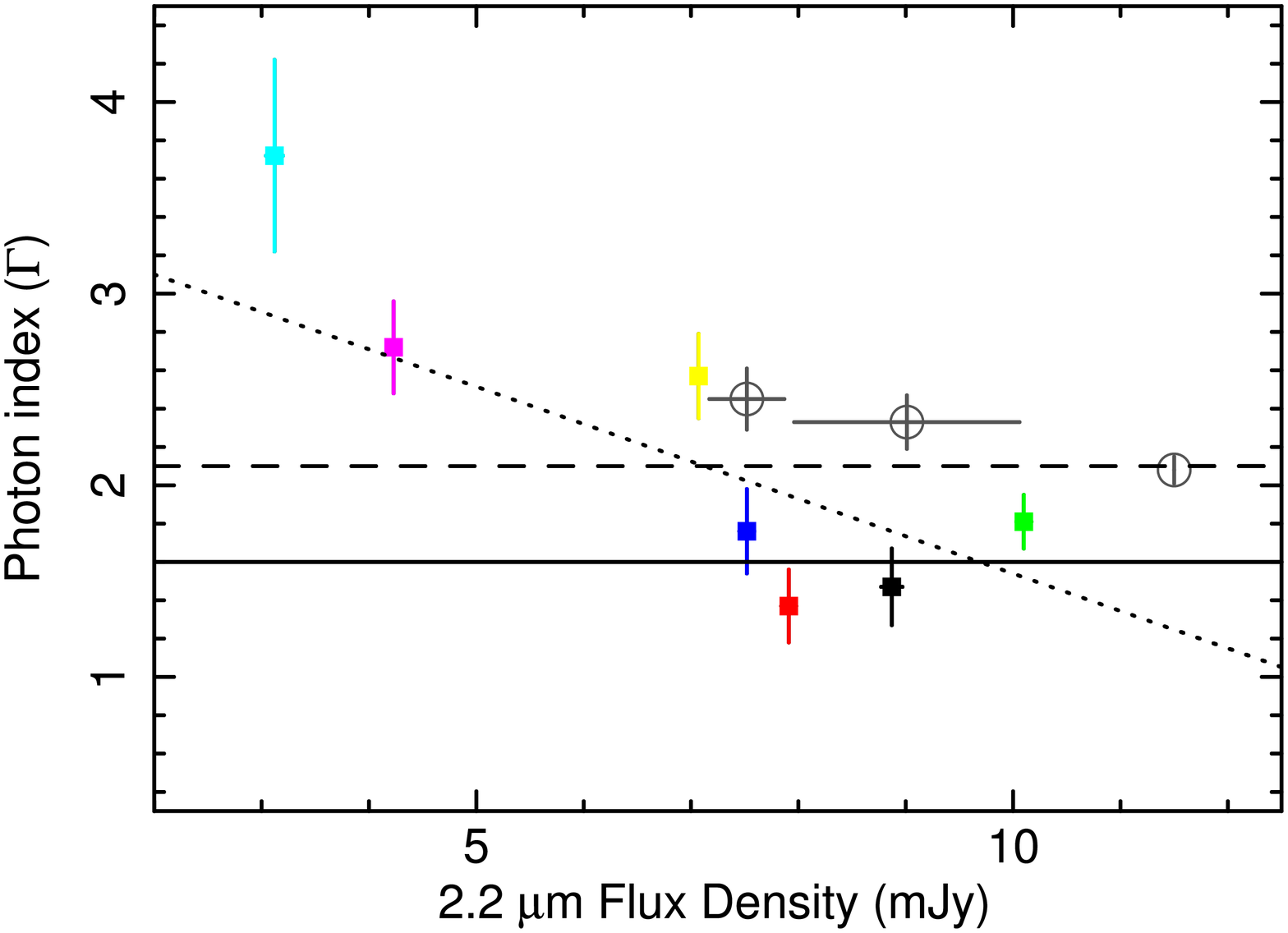}
\caption{{\it (Top panel)} \sinfoni\ spectra fitted with a power-law model 
in the energy band $E\sim0.525-0.608$~eV. 
The color (black, red, green, blue, cyan, magenta and yellow) 
indicates the chronological sequence of the spectra. 
{\it (Bottom panel)} Best fit photon index ($\Gamma_{NIR}$) as a function of 
the $2.2~\mu$m flux density (in mJy units). The NIR photon indexes 
are shown with filled squares, with the same color code as before. 
The empty dark grey circles show the spectral indexes in the 2-10 keV band, 
during the rise, decay and peak of very bright flares. For these points, 
we associate to the flare rise and decay X-ray photon indexes the simultaneous 
NIR fluxes.  For the flare peak, we assume a value of 11.5~mJy. 
The dotted lines show the best fit of the NIR photon indexes with a linear relation. 
The solid line shows the constant photon index typically 
observed at medium-high fluxes (flux density $>7$~mJy; $\Gamma_{NIR}=1.6$; 
Hornstein et al. 2007). The dashed line shows the associated X-ray 
slope, if the spectrum is dominated by synchrotron emission with a cooling 
break $\Gamma_X=\Gamma_{NIR}+0.5$. }
\label{GammaIR}
\end{figure}

We fit all the seven high quality \sinfoni\ spectra (see top panel of 
Fig. \ref{GammaIR}) with a simple power-law model, normalised 
at $2.2~\mu$m ({\sc pegpwrlw}). 
The fit with this simple model provides 
a $\chi^2=96.8$ for 56 dof. The bottom panel of Fig. \ref{GammaIR} 
shows the best fit photon index ($\Gamma_{NIR}$, where $\Gamma=1-\alpha$ 
and $\alpha$ is the spectral index $F_\nu\propto\nu^\alpha$) as a function of the 
flux density (in mJy) at 2.2~$\mu$m. 

During the \sinfoni\ observations the $2.2~\mu$m flux density ranges 
from $\sim3$ to $\sim10$~mJy, spanning the range between 
a classical dim and bright NIR period (Bremer et al. 2011). 
This suggests that this very bright X-ray flare is associated with a very 
bright NIR flux excursion. In agreement with previous results, we observe 
a photon index consistent with $\Gamma_{NIR}=1.6$ above $\sim7$~mJy 
(solid line in Fig. \ref{GammaIR}; Hornstein et al. 2007; Witzel et al. 2014). 
On the other hand, Fig. \ref{GammaIR} also shows steeper NIR spectral 
slopes at low fluxes. We note that steep NIR slopes at low fluxes have 
been already reported (Eisenhauer et al. 2005; Gillessen et al. 2006; 
Bremer et al. 2011), however recent observations by Witzel et al. (2014) 
indicate no spectral steepening at low fluxes. The results of our work appear 
to suggest an evolution of the spectral slope at low fluxes during and after 
this very bright X-ray flare, however higher quality data are necessary 
to finally clarify this trend. 

\subsection{Multi-wavelength spectral evolution during VB3}

We extracted strictly simultaneous \xmm\ and \nustar\ spectra for each 
of the 7 NIR \sinfoni\ spectra (see Fig. \ref{LC}). All are covered by \xmm\ 
and \nustar, apart from spectrum IR4, for which \sgras\ was not visible by 
\nustar\ at that time (due to Earth occultation). The first four (from IR1 
to IR4) of these spectra have been accumulated either when the X-ray 
counterpart of VB3 was visible or in its close proximity and they 
all show bright NIR emission, therefore we present the results of 
the analysis of those "flaring spectra" here. The remaining three, 
associated with faint NIR and X-ray quiescent emission, are investigated 
in the next section (\S \ref{quie}).  

We stress again that during the IR1 spectrum the flare was already 
very bright ($F_{2.2\mu m}\sim9$~mJy) in the NIR band, while only 
upper limits were observed in X-rays (see Fig. \ref{LCp} and \ref{SEDev}). 
Indeed, the X-ray flare started roughly 20 min later, during IR2, and 
peaked just after IR3. 
Bright NIR emission with no X-ray counterpart in the early 
phases of the flare places tight constraints on the PLCool model 
(see \S \ref{discussion}). During IR4 the NIR flux was still high 
($F_{2.2\mu m}\sim7.5$~mJy), while the X-ray flare was about 
to end (Fig. \ref{LCp}). After IR4 the NIR droped significantly and 
the X-ray emission returned to the quiescent level. 
\begin{table} 
\begin{center}
\begin{tabular}{ c c c c c }
\hline
\multicolumn{5}{c}{\bf Simultaneous (600 s) NIR to X-ray spectra during VB3} \\
\hline
\multicolumn{5}{c}{\bf BPL}                                                                               \\
      & $\Gamma_{NIR}$            & $\Gamma_{X}$ & $E_{br}$               & $\chi^2/dof$ \\
      &                                         &                          &  (eV)                             &               \\
IR1 & $1.5\pm0.2$                   & $>2.2$              & 1\ddag                           & 20.8/15 \\
IR2 & $1.4\pm0.2$                   & $3.2\pm0.4$     & $0.16^{+0.20}_{-0.11}$ & 32.8/24 \\
IR3 & $1.8\pm0.2$                   & $2.57\pm0.16$ & $420^{+980}_{-210}$    & 43.8/43 \\
IR4 & $1.8\pm0.2$                   & $2.14\pm0.02$ & 1\ddag                           & 11.4/10 \\
\hline
\multicolumn{5}{c}{\bf PLCool}                                                   \\
      & $\Gamma_{NIR}$&&$E_{br}$                                          & $\chi^2/dof$ \\
      &                        && (eV)                                                &               \\
IR1 & $1.72\pm0.04$&& $0.60\pm0.03$\dag                       & 23.7/15  \\
IR2 & $1.58\pm0.01$&& $0.61\pm0.03$\dag                       & 43.8/25  \\
IR3 & $1.87\pm0.07$&& $530^{+1400}_{-380}$                  & 46.0/44  \\
IR4 & $1.77^{+0.06}_{-0.02}$ && $9.7^{+77}_{-8.0}$             & 11.2/10  \\
\hline
\multicolumn{5}{c}{\bf PLCoolEv} \\
      & $\Gamma_{NIR}$         & $E_c$                    &$E_{br}$                       & $\chi^2/dof$ \\
      &                                      & (keV)                      & (eV)                              &              \\ 
IR1 &$1.48^{+0.25}_{-0.05}$ & $6\times10^{-4}-1$& $0.6-1000$                  & 17.4/14 \\ 
IR2 &$1.5\pm0.2$                  & $3.5^{+0.3}_{-1.1}$& $1^{+2}_{-0.5}$           & 35.6/24 \\ 
IR3 &$1.82\pm0.07$              & $>9$                      & $250^{+780}_{-150}$   & 45.3/43 \\ 
IR4 &$1.77^{+0.06}_{-0.02}$ & $>10$                    & $9.8^{+77}_{-8.3}$       & 11.0/9   \\  
\hline
\end{tabular}
\caption{Best fit parameters of \sgras's emission as fitted during 
each of the (600~s) \sinfoni\ and strictly simultaneous X-ray spectra 
accumulated during the flare VB3. The spectra are fit with both 
the BPL, the PLCool and the PLCoolEv models. 
$\Gamma_{NIR}$ and $\Gamma_X$ indicates the power 
law photon indexes fitting the NIR, the X-ray band, respectively. 
For the PLCool and PLCoolEv models the $\Gamma_{NIR}$ 
indicates the best fit NIR slope, once the total band is fitted with 
the assumption that $\Gamma_X=\Gamma_{NIR}+0.5$.
$E_{br}$ indicates the energy of the cooling break. 
$E_{c}$ indicates the energy of the high-energy cut-off (induced 
by $\gamma_{max}$). \dag The best fit energy of the break falls 
right at the higher edge of the \sinfoni\ energy band. \ddag Unconstrained 
value, therefore fixed to 1~eV.}
\label{TabSEDev}
\end{center}
\end{table} 

\subsection{Is the spectral evolution required?}

We started the time-resolved spectral analysis by testing whether 
the data require any spectral evolution during VB3. 
We therefore simultaneously fitted the multi-wavelength flaring 
spectra (IR1, IR2, IR3 and IR4) with a broken power law model, 
forcing the NIR and X-ray photon indexes and the break energy 
to be constant over time. 
This provides an unacceptable fit ($\chi^2=237.6$ for $104$ dof), 
demonstrating that significant spectral variability is required 
during the flare. The best fit photon indexes are 
$\Gamma_{NIR}=1.71\pm0.09$, $\Gamma_{X}=2.21\pm0.10$, 
with the break at $E_{br}=10^{+150}_{-3}$~eV. We note that, similar 
to what has been found in the analysis of the mean spectrum, the spectral 
steepening is $\Delta\Gamma=0.50\pm0.13$, therefore perfectly 
consistent with $\Delta\Gamma=0.5$. 

We then refitted the spectra with the same model, allowing 
the NIR photon index and the break energy to evolve with time, 
while imposing the X-ray photon index to be 
$\Gamma_{X}=\Gamma_{NIR}+0.5$. 
This provided a significant improvement to the fit ($\Delta\chi^2=97.6$ 
for the addition of 5 new parameters), demonstrating that \sgras's 
spectrum changed shape during VB3. Indeed, we observe best fit 
photon indexes of: $\Gamma_{NIR1}=1.70\pm0.05$, 
$\Gamma_{NIR2}=1.60\pm0.08$, $\Gamma_{NIR3}=1.91\pm0.07$ and 
$\Gamma_{NIR4}=1.81\pm0.13$, while the break is at 
$E_{br1}=0.6\pm0.03$, $E_{br2}=0.9\pm0.03$, 
$E_{br3}=1150^{+1800}_{-800}$ and $E_{br4}=14^{+400}_{-11}$~eV. 
We note that this model can acceptably reproduce the 
data ($\chi^2=140.0$ for 99 dof).

\subsection{Evolution of the BPL model during VB3}

Before considering the PLCool model, where the slopes in NIR 
and X-rays are tied by the relation $\Gamma_{X}=\Gamma_{NIR}+0.5$, 
we fitted each time-resolved multi-wavelength spectrum with the 
phenomenological BPL model (\S \ref{BPL}), where the slopes 
in the NIR and X-ray bands are free to vary (Tab. \ref{TabSEDev}). 

The NIR slope was always well determined ($\Delta\Gamma_{NIR}\sim0.2$; 
see Tab. \ref{TabSEDev} and \ref{IRtime}). 
On the other hand, the presence of either upper limits or low statistics 
prevented us from determining $\Gamma_X$ at the same time of $E_{br}$
in spectra IR1 and IR4 (Tab. \ref{TabSEDev}). We, therefore, "a priori" assumed 
that the breaks in IR1 and IR4 occur at 1~eV (which corresponds to 
$B=30$~G, if interpreted as a cooling break). We then constrained 
the power-law slopes in the X-ray band under this assumption 
($\Gamma_X>2.2$ and $\Gamma_X=2.14\pm0.02$ for IR1 and 
IR4, respectively). Moving the break to 25~eV (corresponding to $B=10$~G) 
the slope would steepen to $\Gamma_X>2.6$ and 
$\Gamma_X=2.4\pm0.1$, respectively. 
For IR2 and IR3, the X-ray data are of good enough quality to have a 
good constraint on the X-ray slope (see Tab. \ref{TabSEDev}). 
The dotted lines in the corresponding panels of Fig. \ref{SEDev} show 
the uncertainties on the X-ray and NIR slopes. 
The BPL model produced an acceptable description of the spectra 
(the surviving residuals are due to intrinsic scatter in the NIR band; 
see Tab. \ref{TabSEDev}). 

\subsection{Evolution of the PLCool model during VB3}
\label{EvPLCool}

We then fitted the spectra with the PLCool model (Fig. \ref{SEDev}; 
Tab. \ref{TabSEDev}). This model reproduces Synchrotron emission 
with a cooling break under the assumption that, at any time, 
$\gamma_{max}>10^6$.

\subsubsection{Successes of the PLCool model}

For IR1, IR3 and IR4 the PLCool model provides a good fit to the data 
of indistinguishable (at 90~\% confidence) quality compared to 
the phenomenological BPL model (Tab. \ref{TabSEDev}). 
The advantage over BPL is that the PLCool model is physically motivated. 

\begin{figure*}
\includegraphics[height=0.49\textwidth,angle=90]{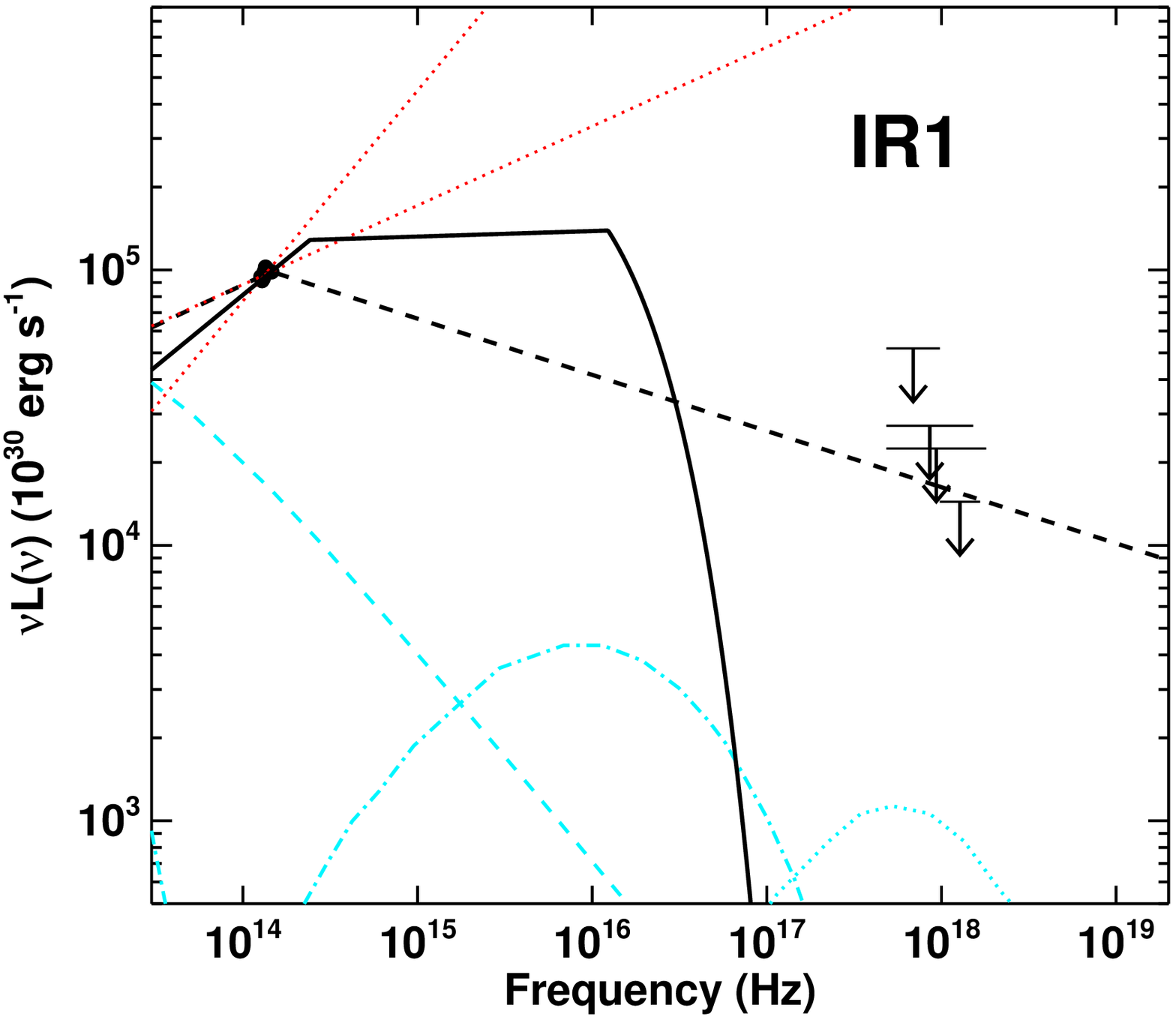}
\includegraphics[height=0.49\textwidth,angle=90]{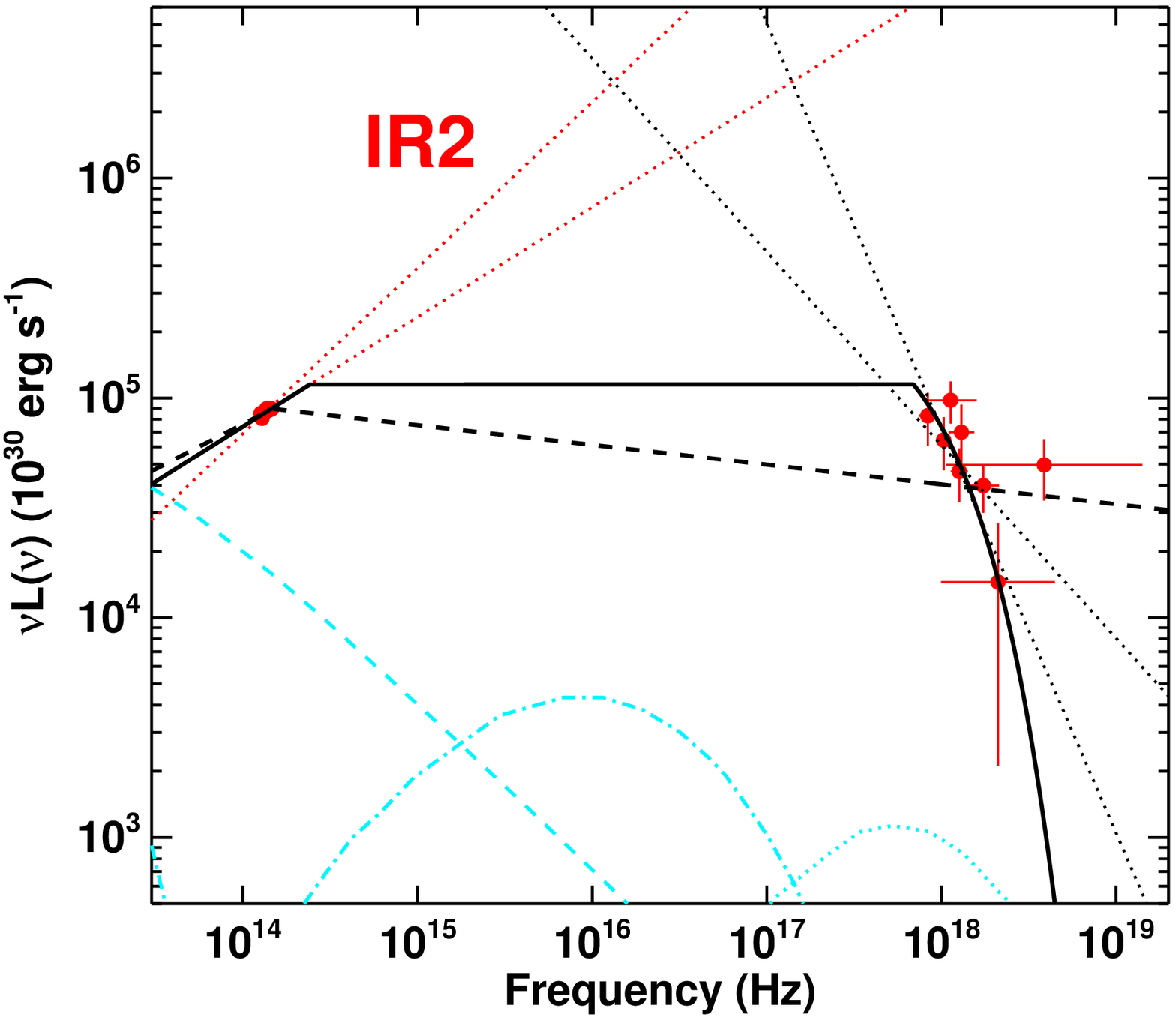}
\includegraphics[height=0.49\textwidth,angle=90]{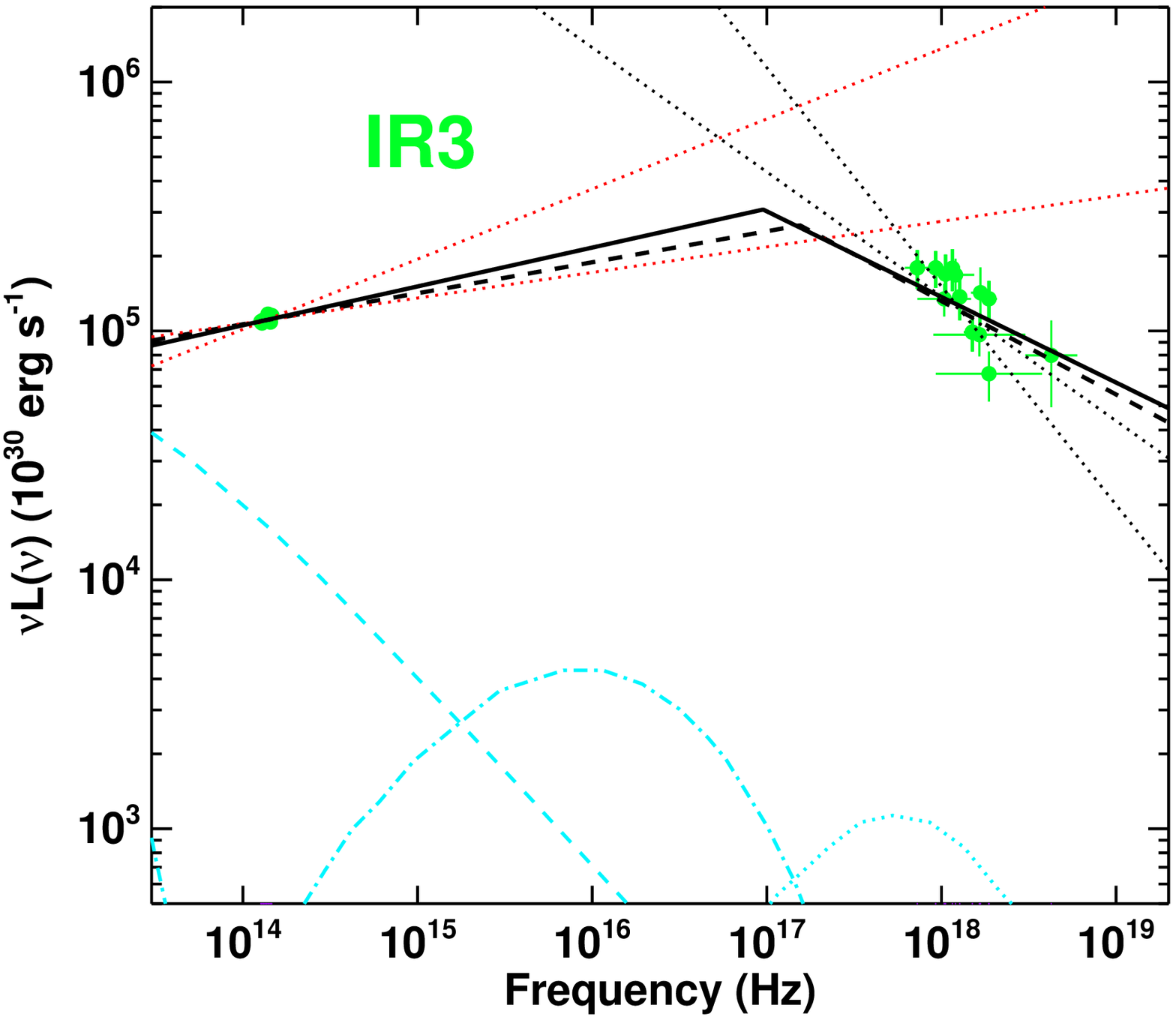}
\includegraphics[height=0.49\textwidth,angle=90]{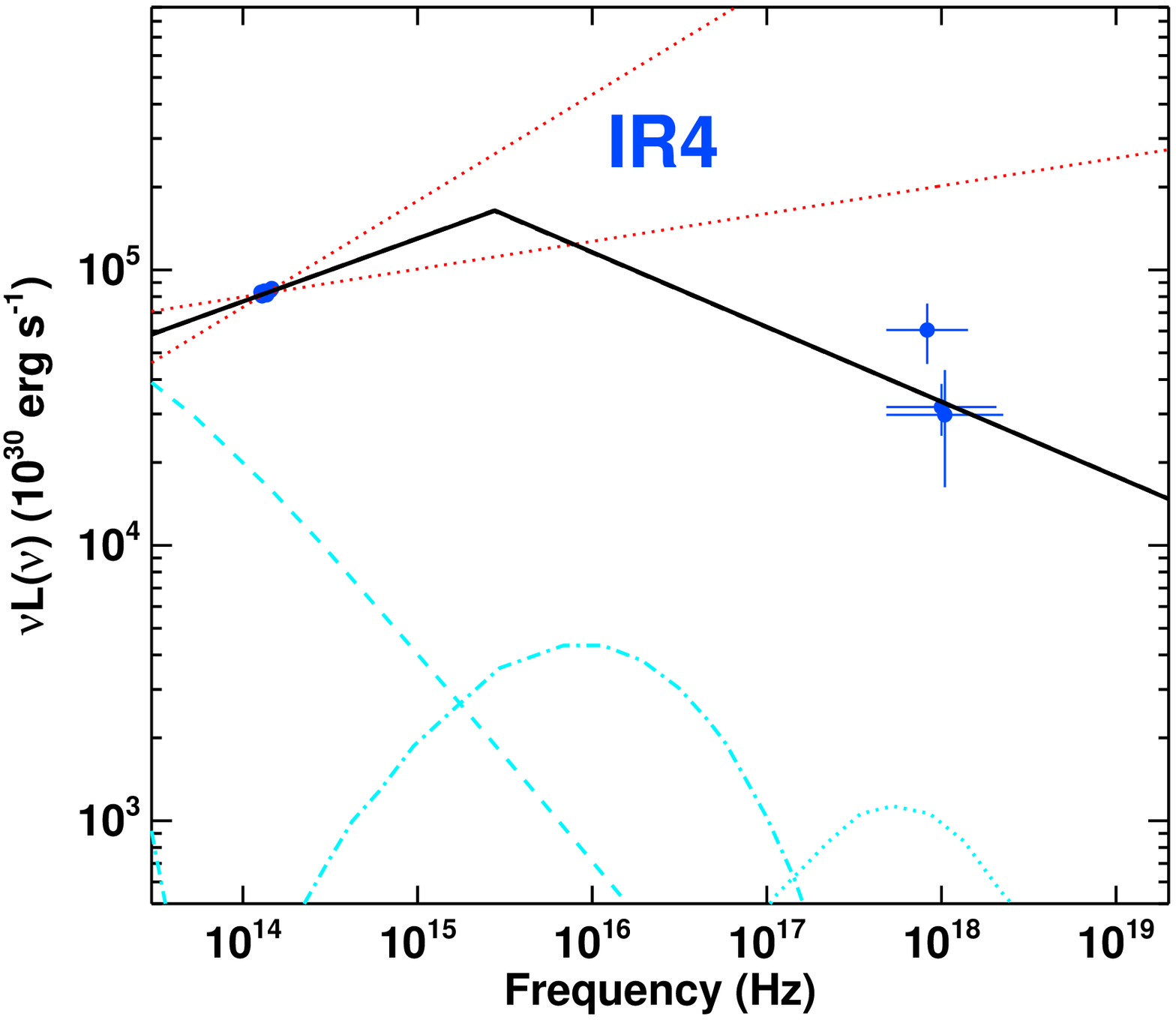}
\caption{Evolution of \sgras's SED during the very bright flare VB3. 
Each panel shows the \sinfoni\ and simultaneous X-ray spectra fitted with 
the PLCool model, during each of the 4 \sinfoni\ spectra integrated 
during the VB3 flare (see Fig. \ref{LCp}). The colour code is the 
same as in Fig. \ref{LCp}, with the temporal sequence: black; red; green and; blue. 
The red and black dotted lines show the uncertainties in the determination 
of the NIR and X-ray power-law slopes, respectively. The X-ray slopes 
are well determined only for the IR2 and IR3 spectra. 
The black dashed lines show the best fit PLCool models, where 
$\Gamma_X=\Gamma_{NIR}+0.5$ is imposed. For IR2 the observed X-ray 
slope is inconsistent with the predictions of the PLCool model. 
For both IR1 and IR2 the cooling break is suspiciously pegged in 
the NIR band. The black solid lines show the best fit PLCoolEv models. 
During IR1 both the cooling break and the cut-off have large uncertainties, 
but are constrained to lie within few $10^{14}<\nu<10^{18}$~Hz. During IR2 
the cut-off is in the X-ray band. 
From IR2 to IR3 the cooling break evolves to higher energies and then 
back to lower energies during IR4. 
\sgras\ is undetected in X-rays during observation IR1. 
Such as in Fig. \ref{PLCool} both data and models are de-absorbed and 
corrected for the effects of the dust scattering halo. 
For a description of the other lines see Fig. \ref{SED}.}
\label{SEDev}
\end{figure*}

We observed that for all spectra (from IR1 to IR4) the NIR spectra 
are flat and consistent with being 
constant (e.g. $\Gamma_{NIR}\sim1.6$, see Fig. \ref{GammaIR}) 
before and during the full duration of the X-ray flare. This is in line 
with the values typically observed during bright NIR flux excursions 
(Hornstein et al. 2007)\footnote{More observations are needed to 
confirm the tentative hint for a steeper NIR slope in the early phase 
of the NIR flare (IR1 and IR2 compared to IR3 and IR4; see 
\ref{TabSEDev}).}. We also noted that at the peak of the X-ray flare, 
when the constraints are best, the X-ray slope is steeper than 
the simultaneous NIR one by $\Delta\Gamma\sim0.5$, 
consistent with the one expected by the PLCool model (see 
Fig. \ref{SEDev} and Tab. \ref{TabSEDev}). 

\begin{figure}
\includegraphics[height=0.49\textwidth,angle=-90]{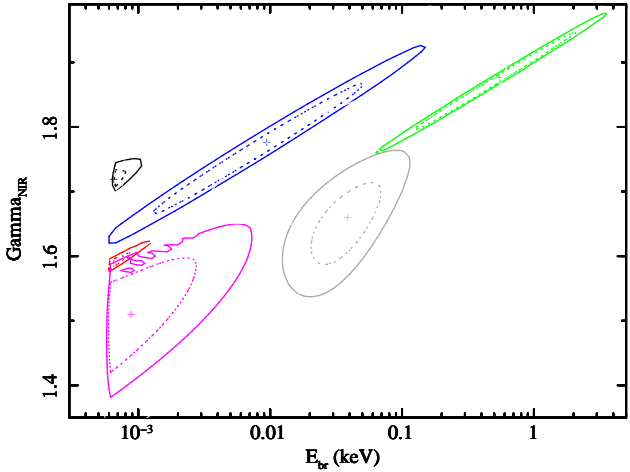}
\caption{The black, red, green, blue and grey lines show confidence 
contours of the uncertainty on $E_{br}$ and $\Gamma_{IR}$ for IR1, 
IR2, IR3, IR4 and the mean spectrum once fitted with the PLCool model, 
respectively. The dotted and solid lines show the 68 and 90~\% confidence 
contours, respectively. During IR1 and IR2 the cooling break pegs 
at its lowest value, being located just above the NIR band. 
Once fitted with the PLCoolEv model, the confidence contours remain 
unchanged for IR3, IR4 and the mean spectrum, because during 
these intervals the high energy cut off is at very high energy. 
The magenta lines show the confidence contours during IR2, 
when the cut off is observed in the X-ray band. During IR1, the cut off 
falls between the NIR and X-ray band, therefore the location of 
the cooling break is unconstrained. }  
\label{Cont}
\end{figure}

Blindly applying the PLCool model to all time resolved spectra 
(it might be incorrect to apply the PLCool model when no X-ray 
emission is detected), we observed a significant evolution 
of the energy of the cooling break, that implies 
(under the assumption of a constant escape time)
a variation of the strength of the magnetic field (Tab. \ref{TabSEDev}).  
The black, red, green, blue and grey dotted and solid lines 
in Fig. \ref{Cont} show the 68 and 90~\% confidence contours 
of the uncertainty on $E_{br}$ and $\Gamma_{IR}$ for IR1, IR2, IR3, 
IR4 and the mean spectrum, respectively. We note that a highly 
significant evolution of the cooling break is observed. 
Indeed, during both IR1 and IR2, the break appears to be at very 
low energy, corresponding to a magnetic field of the order of 
$B\sim35$~G. While the energy of the break is significantly 
higher during IR3, indicating that the magnetic field had significantly 
reduced around the peak of the X-ray flare ($B=3.8^{+2.0}_{-1.3}$~G). 
The energy of the break then drops again in the decreasing flank of 
the X-ray flare to a value of $E_{br}=9.7^{+77}_{-8.0}$~eV, 
corresponding to an increase in the strength of the magnetic 
field ($B=14.3^{+11.3}_{-7.4}$~G). 

\subsubsection{Difficulties of the PLCool model}
\label{PLCoolEv}

As we have outlined, the PLCool model (which assumes 
$\gamma_{max}>10^6$ at all times) presented many successes. 
However, we also point out here three severe weaknesses 
that will be discussed further in the discussion section: 
i) twice out of four times the cooling break is observed to peg 
just above the NIR band ($E\sim0.6$~eV). This appears as a rather 
unlikely possibility; ii) during IR1, bright and flat 
($\Gamma=1.48\pm0.2$) NIR emission is associated to no 
enhanced X-ray emission 
($F_{3-10~keV}<2.3\times10^{-12}$~erg~cm$^{-2}$~s$^{-1}$). 
This is hard to reconcile with the PLCool model that, in order to 
fit this spectrum, pushes the best fit NIR photon index 
to $\Gamma_{NIR}=1.72\pm0.04$; iii) the very steep 
X-ray spectrum during IR2 ($\Gamma_X=3.2\pm0.4$) implies a 
spectral steepening incompatible with the PLCool model 
($\Delta\Gamma=1.8\pm0.4$, instead of $\Delta\Gamma=0.5$). 

\subsection{Can the TSSC model fit the IR2 spectrum?}

In theory SSC models, with a thermal distribution, can produce fairly 
steep spectral shapes at high energies. Therefore, although the TSSC 
model produced unreasonable parameter values when applied to the mean VB3 
spectrum, we checked whether TSSC might be the dominant radiative 
mechanism during peculiar and short duration intervals, such as IR2. 

The best fit TSSC model was significantly worse than the PLCool model, 
despite having 
two more free parameters ($\chi^2=60.1$ for 23 dof). Indeed, the model 
failed to produce a better fit because it was mainly constrained by the flat 
X-ray photon index produced by the TSSC model. 
In addition, we noted that 
the fit of IR2 leaded to unreasonable best fit parameters, similar to the 
ones fitting the mean VB3 spectrum. 
Indeed, we observed: $Log(B)=3.55$, $\theta_E=32.6$, $Log(N_e)=39.3$ 
and $Log(R_F/R_S)=-3.0$. Once again the magnetic field strength 
appears, unreasonably large, the size of the source unreasonably 
small, and the source density many orders of magnitude higher than 
expected. 

\subsection{Synchrotron emission with cooling break and evolving 
$\gamma_{max}$ (PLCoolEv)}
\label{Evgamma}

When we considered the PLCool model, we "a priori" used 
the assumption that at any time the source can accelerate particles 
to very high energies $\gamma_{max}>10^6$. This implies 
that an engine, able to accelerate electrons to $\gamma_{max}>10^6$, 
is created on a negligible time-scale, at the start of the NIR flare. 
If so, the PLCool model can be applied to the entire duration of 
the flare (as we performed in \S \ref{PLCoolEv}). 

On the other hand, if the engine has a size of a few Schwarzschild 
radii, its light (or Alfven speed) crossing time would be of the order 
of a hundred seconds, comparable to the time-scales of the flares 
under investigation here. 
Thus, it might be possible that the creation and destruction of the engine 
occurs on a similar time-scale to the flares and that the engine is 
initially not powerful enough to accelerate 
particles to $\gamma_{max}>10^6$. Based on these considerations, 
we introduce a new phenomenological model dubbed PLCoolEv, 
by adding to the PLCool model the freedom of having a variable value of 
$\gamma_{max}$. We performed this by adding a high energy 
cut off to the PLCool model (reproduced by the {\sc highecut} component 
in {\sc XSpec}). Indeed, the PLCoolEv model assumes that 
$\gamma_{max}$ evolves during the flare. A low value of 
$\gamma_{max}$ (e.g., $\gamma_{max}<<10^6$) would imply 
that no electrons are accelerated to such high energies to produce 
X-ray photons\footnote{Assuming that all emission is radiated at 
0.29 times the critical frequency, it follows that: 
$\nu_c\sim2.5\times10^{19}$~Hz $(\gamma_{max}/10^6)^2$ $(B/20$~G$)$ 
(Longair 2011, equation 8.127), where $\nu_c$ is the frequency 
associated to the high energy cut off.}. As a result, the emitted spectrum 
would show a high energy cut off ($E_c$) at an energy related to 
$\gamma_{max}$ and lower than the X-ray band. 
As in the PLCool model (with free parameters: $B$, $p$ and 
normalisation), this model is characterised by the cooling 
break $E_{br}$, linked to the strength of the magnetic field, 
plus a cut-off at high-energy $E_c$ (induced 
by $\gamma_{max}$). 

We assumed an exponential shape above the cut off energy. 
We noted that such shape is constrained only by the IR2 spectrum 
and it appears steeper ($\Gamma_X=3.2\pm0.4$) 
than the simultaneous and cooled NIR slope (see Tab. 6). 
Therefore any high energy slope steeper than $\Gamma_X\sim3$ could 
reproduce the data. We point out that either an exponential or sub-exponential 
slope can equally fit the data. We also note that most likely the electron
distribution will not cut abruptly at $\gamma_{max}$, therefore it is expected 
that the cut off will be further broadened. Despite we could not constrain 
whether the break is broad, we fixed the shape of the high energy cut off 
such that the e-folding energy of the exponential cut off is equal to the cut 
off energy. 

The PLCoolEv model provides an excellent representation 
of the multi-wavelength spectrum at all times during the flare 
(see Tab. \ref{TabSEDev} and Fig. \ref{SEDev}). 
It produces either superior fits compared to PLCool model 
(in particular for IR2), or of comparable statistical quality 
to the BPL parametrisation and it is physically motivated. 

Figure \ref{Cont} shows the confidence contours projected 
over the $E_{br}$ versus $\Gamma_{NIR}$ plane for 
the PLCoolEv model. The high-energy cut-off is constrained to be 
at $7\times10^{-4}<E_c<1$~keV in the IR1 spectrum, before the start 
of the X-ray flare (not shown in Fig. \ref{Cont}). In the PLCool model, 
the steep photon index observed in X-ray during IR2 is the result of 
the evolution of the high energy cut-off, which at that time was detected 
in the X-ray band at $E_c=3.5^{+0.3}_{-1.1}$~keV. As a consequence 
of this, the cooling break is not pegged anymore at $E_{br}=0.6$~eV, 
instead it spans a larger range of reasonable cooling break energies.
Additionally, a flatter NIR slope is allowed.
During IR3, IR4 and the mean spectrum, the high energy cut-off 
was at energies higher than the observed X-ray band ($E_c\gg10$~keV), 
consistent with the assumptions of the PLCool model (indeed, we obtained 
similar results). According to the PLCoolEv model, 
during the early phase of VB3, the high energy cut-off was evolving 
and it was located between the NIR and X-ray band. It was caught 
within the X-ray band during IR2 and it was at very high energy 
at the X-ray peak (and during IR4). 

\subsubsection{Evolution of the magnetic field (assuming a constant escape time)}
\label{SecEvB}

In this section we interpret the derived evolution of the energy of 
the cooling break, as being uniquely due to the variation of 
the magnetic field of the source (e.g., assuming no variation of the 
escape time). 

\begin{figure}
\includegraphics[height=0.35\textwidth,angle=0]{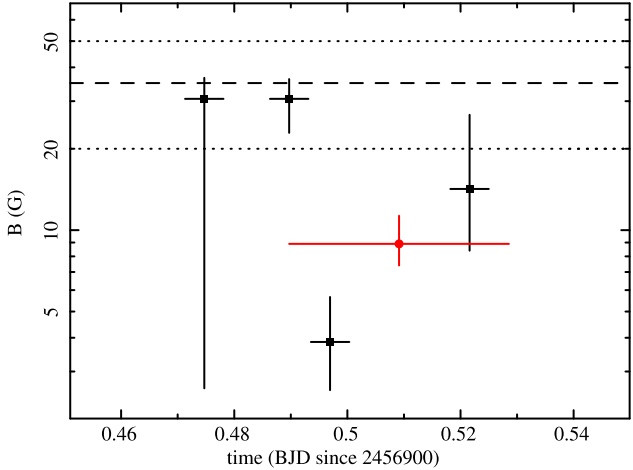}
\caption{Evolution of the strength of the magnetic 
field (in Gauss) during and after the flare VB3, in the 
PLCoolEv model. The black square show the 
magnetic field strength during IR1 to IR4. The red circle shows 
the measurement of the average magnetic field during the entire 
duration of the X-ray flare VB3 (under the assumption of a constant 
escape time of $t_{esc}=300$~s). The dotted lines indicate the typical 
range of magnetic field strengths during quiescence ($B\sim20-50$~G). 
The dashed line shows the magnetic field strength ($B\sim36$~G) 
corresponding to a cooling break within the narrow \sinfoni\ band. 
The flat NIR slopes observed at all times during the flare suggest 
$B<36$~G, while the steep NIR slope observed after the VB3 flare 
(during IR5, IR6 and IR7) suggest $B>36$~G. The error bars 
correspond to the 1-sigma uncertainties as derived from the confidence 
contours shown in Fig. \ref{Cont}. No evidence for a cut off at or 
below the X-ray band is observed during IR4, therefore the associated 
measurement is valid under the assumption that the cut off is at energies 
higher then the X-ray band. Would this assumption be invalid, 
such constraint should be considered as a upper limit. }
\label{EvolPLCool}
\end{figure}

Figure \ref{EvolPLCool} shows the light curve of the evolution 
of the magnetic field intensity during the flare. 
The flat NIR slope observed at all times indicated 
that the break has to be, at higher frequency compared 
to the \sinfoni\ band, corresponding to $B<36$~G. 
In particular, during IR2 the cooling break is observed at 
$E_{br}=1^{+2}_{-0.5}$~eV, corresponding to $B=30\pm8$~G. 
The values derived by fitting the IR1 spectrum are consistent 
with this value, however the degeneracy between the energy of 
the cooling break and of the high energy cut-off led to large 
uncertainties on the magnetic field strength. 
We note that a value of $B=30\pm8$~G is fully consistent with 
the magnetic field present within the central ten Schwarzschild 
radii and generating the steady emission of \sgras. 
During IR3, close to the peak of the X-ray flare, the magnetic 
field is observed to be $B=4.8\pm1.7$~G. Interestingly, 
the magnetic field varied by a factor of $>6$ in less 
than $\sim650$~s. During IR4, we assumed that the cut off 
is located at energies higher than the X-ray band (indeed no evidence 
for a cut off at or below the X-ray band is observed). Under this 
assumption\footnote{ Would, during IR4, the cut off be located 
in the X-ray band or below, then the current constraints on the 
energy of the cooling break should be considered only as upper limits. 
If this is indeed the case during IR4, then weaker magnetic fields would 
be allowed and the data point in Fig. \ref{EvolPLCool} should be 
considered as an upper limit. }, we observe that after the X-ray peak 
and towards 
the end of the X-ray flare, the magnetic field was measured 
to rise again to values $B=14.3^{+12.3}_{-7.0}$~G. The red point in Fig. 
\ref{EvolPLCool} shows the magnetic field strength derived from 
the fit of the mean spectrum of VB3. As expected, the average 
magnetic field value during the flare ($B=8.8^{+5.0}_{-3.0}$~G) 
was intermediate between IR2, IR3 and IR4 and it was significantly 
smaller than the one derived during quiescence. 

\subsubsection{Evolution of the escape velocity (constant magnetic field)}

The results obtained in section \ref{SecEvB} are valid under 
the assumption that the synchrotron escape time is constant over the 
entire duration of the flare. However, it is not a priori set that the escape 
time has to remain constant over time. Therefore, we now investigate 
the hypothesis that the escape time evolves, while assuming 
a constant magnetic field ($B=30$~G).  
If so, the energies of the break frequencies observed during IR2, 
IR3 and IR4 ($E_{br}=1^{+2}_{-0.5}$, $250^{+780}_{-150}$
and $9.8^{+77}_{-8.3}$~eV) correspond to an escape velocity of 
$t_{esc1}=310\pm130$~s, $t_{esc2}=20\pm10$~s and 
$t_{esc3}=100^{+150}_{-70}$~s. Therefore, the escape time would 
drop by a factor of $\sim16$ in $\sim600$~s, to then increase again. 

The escape time is likely related to the source size and/or to the position 
of the source onto the accretion disc. For example, Dodds-Eden et al. (2009) 
assume that the escape time is comparable to the dynamical time at a given 
radius in an accretion disc ($t_{dyn}$): 
$t_{esc}\sim t_{dyn} = \sqrt{R^3/2GM_{BH}}$, where $R$ is the source 
position within the accretion disc, $G$ is the gravitational constant and 
$M_{BH}$ is the black hole mass. 
We note that, under this assumption, the escape time assumed throughout 
the paper ($t_{esc}=300$~s) corresponds to a reasonable radial position of 
$\sim3.5$~R$_S$ from the BH. 

\

In summary, the PLCoolEv can adequately fit not only the mean properties 
of the VB3 flare but also its evolution. The major weaknesses of the PLCool 
model are solved by allowing the high energy cut-off ($\gamma_{max}$) 
to evolve during the flare. 

We note that the evolution of the cooling break appears more 
likely induced by a variation of the magnetic field that drops its intensity 
by discharging magnetic energy density into particle acceleration 
and then rises again to its average value, compared to a variation 
of the escape velocity. Indeed, in the latter scenario it would naively 
be expected that the energy release produced by the source would 
make the source size expand with time, instead of contracting. 
However, we point out that these considerations are not conclusive. 
Indeed, because of the limitations of our simplified single 
zone models, we can not discriminate between a pure magnetic field 
evolution or a pure escape time evolution (or a combination of both).  

\section{Emission after VB3 (X-ray quiescence)}
\label{quie}

IR5 and IR6 have been accumulated after the end of VB3 when 
only upper limits are observed in X-rays and the NIR flux 
($F_{2.2\mu m}\leq4.5$~mJy) corresponds to the faintest fluxes 
of \sgras, detected so far (e.g. Dodds-Eden et al. 2010). 
These time intervals appear similar to classical quiescent periods. 
During IR7 a re-brightening is observed in NIR, associated with a hint 
for an excess in the X-ray band (Fig. \ref{LCp}). Though the NIR 
flux is relatively high ($F_{2.2\mu m}\sim7$~mJy), the NIR spectral 
slope appears steeper than during the flare and fully consistent with 
the value observed during NIR quiescence. The sparse NIR light curve 
as well as the low significance of the X-ray excess do not allow us 
to clarify whether the emission during IR7 is produced by a faint flare, 
with associated feeble X-ray emission or it is just a NIR fluctuation 
characteristic of a red noise process, commonly occurring during 
X-ray quiescence. 

We observed that the NIR spectra steepened $\sim15$~min 
after the end of the X-ray flare (see Tab. \ref{IRtime}). 
The steepening was so large 
($\Delta\Gamma_{NIR}\sim1-2$) that in all cases (from IR5 to IR7) 
the extrapolation of the steep power-law observed in the 
\sinfoni\ band (Tab. \ref{IRtime}) was consistent with the X-ray 
upper limits. Figure \ref{GammaIR} shows 
that at medium-high NIR fluxes ($F_{2.2~\mu m}>7$~mJy) the photon 
index is consistent with a constant value of $\Gamma_{NIR}=1.6$. 
This implies an electron distribution index of $p\sim2.2$. 
We note that in the PLCool (and PLCoolEv) model, as long as 
$B>35-40$~G, the cooling break would move to frequencies lower 
than the \sinfoni\ band, inducing a steepening of the observed photon 
index by $\Delta\Gamma_{NIR}=0.5$. 
However, this steepening appears too small to reproduce the full 
extent of the observed photon index variation. 
We remind the reader that accurate photon index determination 
at low NIR fluxes are challenging. Therefore, we leave to future dedicated 
studies to establish the full extent and the reliability of the NIR spectral 
steepening at low NIR fluxes. If future data confirms the presence 
of such steep NIR slopes after very bright flares, then this radiation 
might be associated with thermal Synchrotron emission from electrons 
transiently heated during VB3. 
Indeed, we note that, for a magnetic field strength of $B\sim30$~G, 
the cooling time of NIR synchrotron electrons is of the order of 
$\sim500$~s, shorter, but comparable, to the time interval between
the end of IR4 and the start of IR5 (see Fig. \ref{LCp} and Tab. \ref{IRtime}). 

\section{Does a slow evolvution of $\gamma_{max}$ agree with 
the evolution of bright X-ray flares?}
\label{SpEvol}

The detailed investigation of the X-ray and NIR emission during VB3 
indicates the PLCoolEv as the favourite model (see \S 5 and 6). 
The essential component that distinguish the PLCoolEv model from 
the simpler PLCool model is the evolution of the cut-off ($\gamma_{max}$).
In particular, we suggest that the evolution of $\gamma_{max}$ might 
be relatively slow, spanning the range from optical-NIR to X-rays and 
beyond on macroscopic time-scales ($\sim10^2-10^3$~s). 
We also note that the X-ray band has significant extensions in 
frequency, spanning over a decade in frequency. 
Therefore, should the PLCoolEv model be correct and should the 
behaviour observed during the very bright flare VB3 be universal, 
then this model would predict an energy dependent evolution of 
X-ray flares that might be tested with archival data of other bright 
X-ray flares. 

Indeed, it is expected that the passage of the cut-off (induced 
by $\gamma_{max}$) through the X-ray band would induce slightly 
shorter flares at higher energies as well as steeper spectral slopes 
at the start and end of the X-ray flare. 
Clearly the full extent of these effects can not be predicted, because 
it depends on how rapidly the cut-off spans the X-ray band, but 
we investigated whether we can exclude that such evolution is present 
during bright X-ray flares. 
Indeed, despite the fact that at present there are only few bright and 
very bright flares with multi-wavelength coverage and only one (VB3) 
with simultaneous NIR and X-ray spectra, the \xmm\ archive contains 
several bright flares suitable for studying the spectral evolution 
in the X-ray band\footnote{We do not consider bright flares 
detected by \chandra, because the vast majority of those are affected
by strong pile-up, significantly distorting the spectral shape. }. 

\subsection{Time dependence of X-ray spectra of very bright flares}

To follow the evolution of \sgras's X-ray emission during very bright 
X-ray flares, we consider here all bright and very bright flares 
observed by \xmm\ (see Tab. \ref{obsid}; Ponti et al. 2015a). 
For each of these flares we extract three spectra, one during the rise, 
one at the peak and one during the decay (see Tab. 
\ref{obsid})\footnote{For flares VB1 and VB2 we chose three intervals 
of equal duration, while for flare VB3, the extraction of the spectra 
during flare rise, peak and decay are chosen in order to optimise the 
coverage of the \sinfoni\ spectra (see Tab. \ref{obsid}, \ref{IRtime} and 
Fig. \ref{LCp}).}. 
We fit the spectra from both the EPIC-pn and MOS data during the peak 
of VB1, VB2 and VB3 with the absorbed power-law model. 
The observed spectral indices 
at peak are consistent between the different flares; therefore we assume 
the same value and redo the fit to obtain a best fit value of $\Gamma=2.08\pm0.11$
($\pm0.07$ at 1 $\sigma$). We then repeat this exercise fitting the spectra
during both the flare rise and decay, obtaining $\Gamma=2.33\pm0.23$ 
and $\Gamma=2.45\pm0.25$, respectively. Again, we observe consistent 
values for the rises and decays of different flares. Therefore we assume 
the same spectral index during both flanks of the flares, obtaining a 
best fit value of $\Gamma=2.36\pm0.15$ ($\pm0.09$ at 1 $\sigma$), 
slightly steeper ($\sim2.4 \sigma$ significance) than the spectral index 
observed at peak. We conclude that the spectra of very bright flares 
provide hints for (or at least are not in disagreement with) an evolution 
of the order of $\Delta\Gamma\sim0.3$ between the peak and 
the flanks of the flares. This behaviour is reminiscent of what was 
observed during the evolution of VB3. Indeed, during the early phases 
of the X-ray emission (IR2), the X-ray spectrum was steeper than at 
the peak of the X-ray emission (IR3; \S \ref{SecSpEvol}), most 
likely because of the evolution of $\gamma_{max}$.

\subsection{Colour dependence of bright X-ray flares}

\begin{figure}
\includegraphics[height=0.5\textwidth,angle=90]{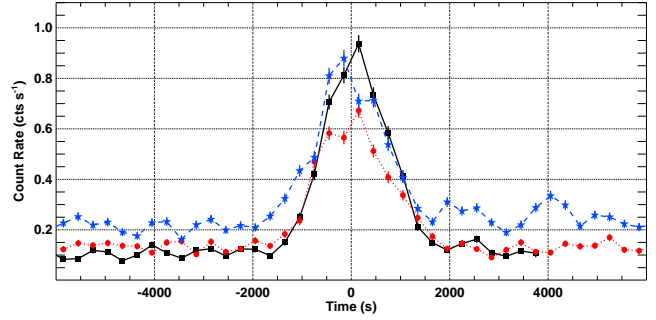}
\caption{Light curves, in the 2-10~keV band, of the three brightest \xmm\ 
flares, VB1, VB2 and VB3 are shown with back squares, red circles and 
blue stars, respectively. 
The light curves of the three very bright flares observed by \xmm\ show
very similar time evolution and comparable duration. These light curves
are the result of the sum of the data from EPIC-pn and MOS.
For display purposes, we shifted the time axis aligning the peak of 
the Gaussian best fitting the flare profile (see Tab. \ref{modLC}). }
\label{LC}
\end{figure}

Figure \ref{LC} shows the 2-10 keV band light curves of the three 
very bright flares observed by \xmm\ (Ponti et al. 2015a). We combined 
the light curves from the three EPIC cameras. 
Figure \ref{LC} shows a remarkable similarity in the evolution of these 
flares, suggesting an analogous origin.  
We fit each of the light curves with a model composed by a constant 
plus a Gaussian profile (to fit the flare)\footnote{The variation of the non-flare 
emission during both observations taken on 2007-04-03 and 
2014-08-31 was produced by the contribution from the magnetar \sgr\ 
(at the level of $\sim50$ \% to the total observed quiescent 
flux on 2014-08-30; see also Ponti et al. 2015a) and from the very 
bright source \axj\ (Ponti et al. 2015b). }. 

\begin{figure}
\includegraphics[height=0.47\textwidth,angle=-90]{length4.ps}
\includegraphics[height=0.33\textwidth,angle=0]{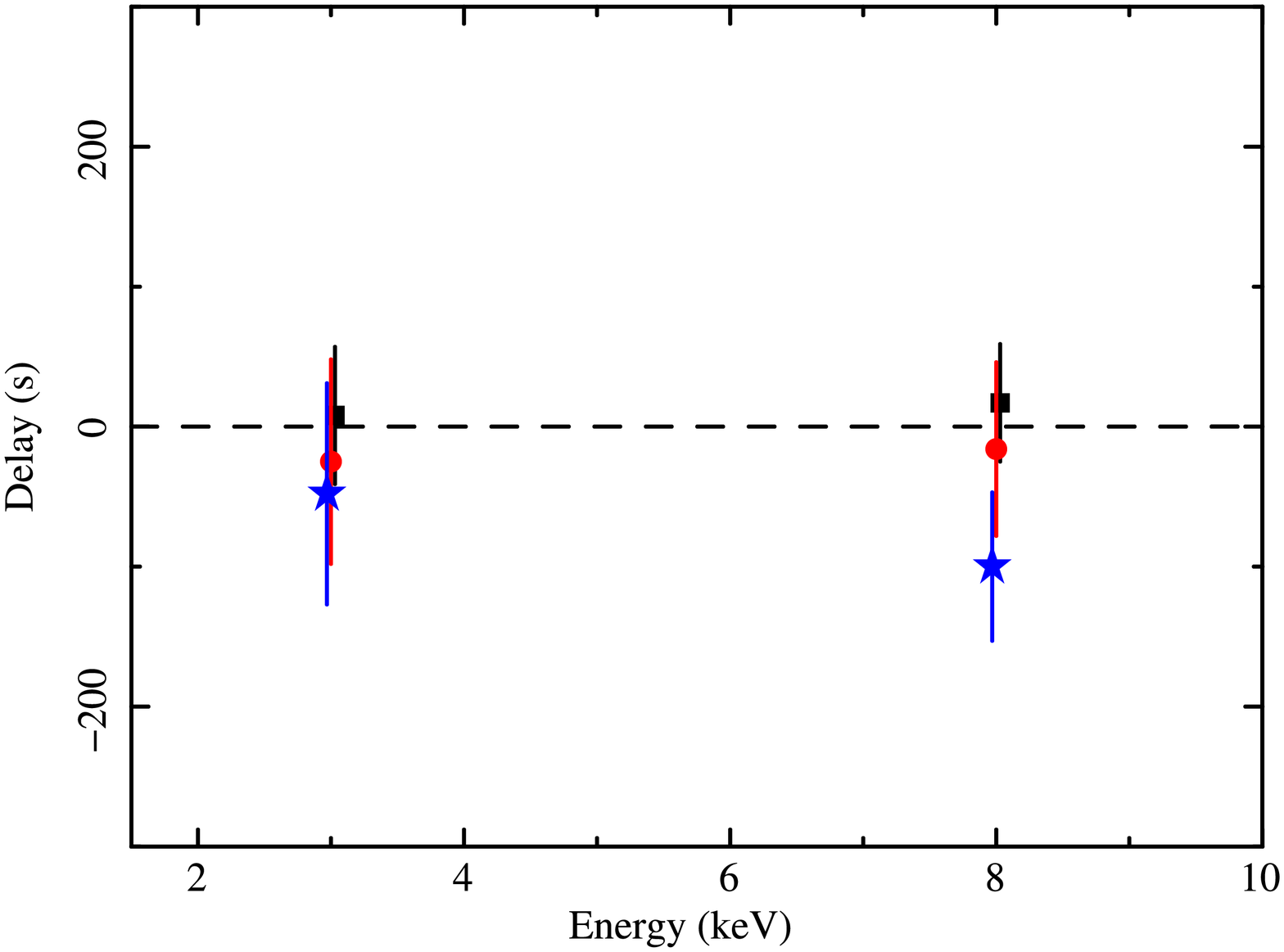}
\caption{{\it (Top panel)} Flare duration (FWHM) at various energies 
of the three brightest \xmm\ flares, VB1, VB2, VB3 and of the combined 
\xmm\ flare light curve are shown with back squares, red circles, blue stars 
and green triangles, respectively. The flare duration are computed in the 
2-4, 4-6 and 6-10~keV energy bands (the points are slightly shifted for 
display purposes). Flares are typically shorter at higher energies. 
We also show the duration of VB2 (as observed in the L' band, 
see \S \ref{NIR}) and the lower limit on the duration of VB3 (as observed 
with \sinfoni). The flare durations in the NIR band connect with the extension 
of the trend observed in X-rays. 
{\it (Bottom panel)} Peak occurrence delay, between different energies 
(same energy bands as above) for the three very bright flares observed 
by \xmm. The delays are computed as the best fit peak value of the 
Gaussian at each energy minus the same value observed in the 4-6 keV 
band. Colour code as before. }
\label{LCEne}
\end{figure}

To investigate possible dependences of the X-ray flares on energy, we 
extracted the X-ray light curves of the three very bright \xmm\ flares 
in the 2-4, 4-6 and 6-10 keV energy bands. For each flare, we fit 
the light curves with a constant plus a Gaussian profile, to characterise 
the flare shape (see Table \ref{modLC}). 

The top and bottom panels of Fig. \ref{LCEne} show the best fit 
flare duration ($FWHM$) and delay as a function of energy. 
For each energy we report with black squares, red circles and blue stars 
the values obtained for the flares VB1, VB2 and VB3, respectively. 
In particular, we show the width of the best fit Gaussians as a proxy 
for the flare duration and the delay is defined as the peak time of the 
Gaussian at a given energy minus the peak time in the 4-6~keV band. 

\begin{figure}
\includegraphics[width=0.23\textwidth,angle=90]{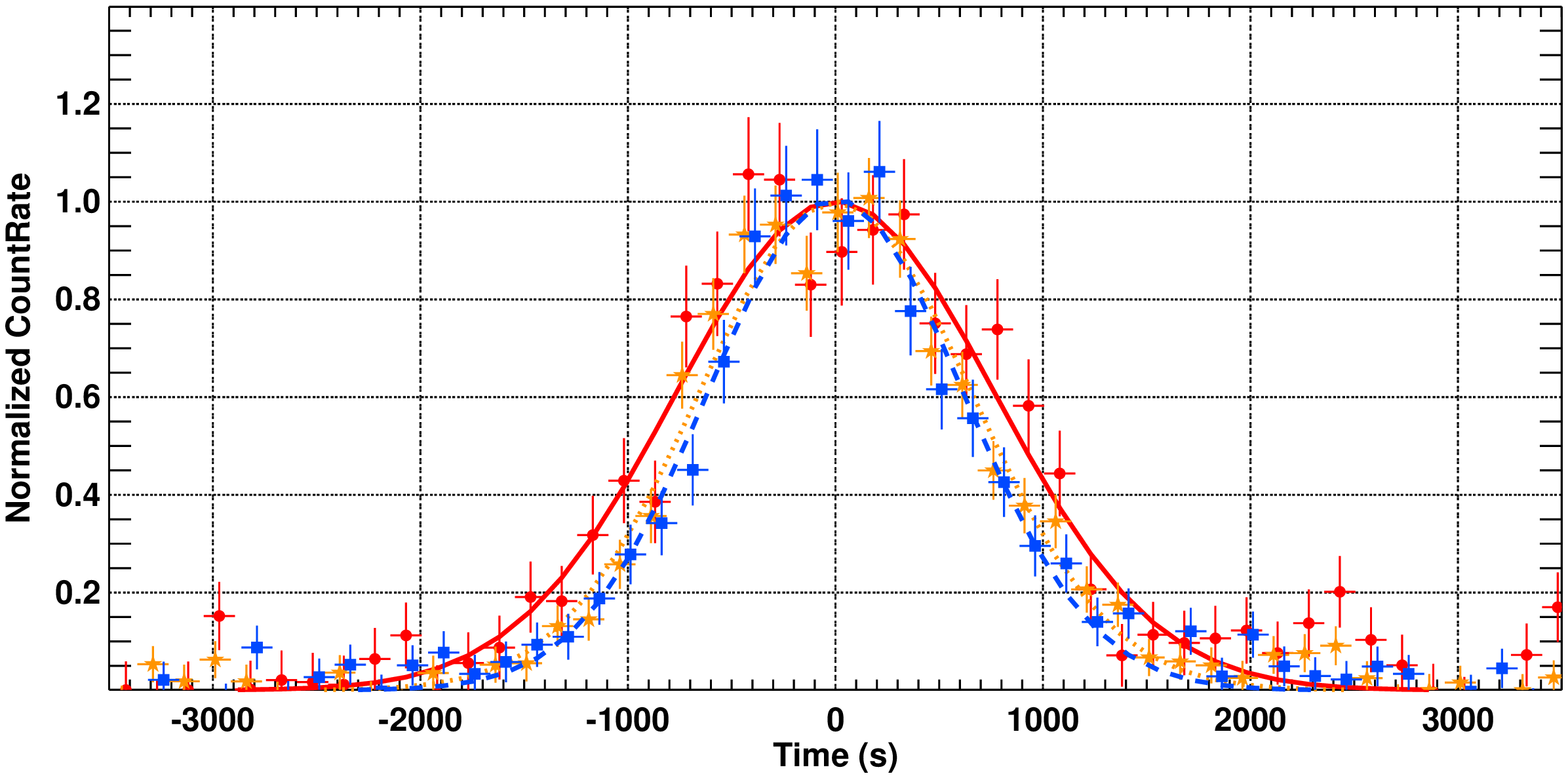}
\includegraphics[width=0.23\textwidth,angle=90]{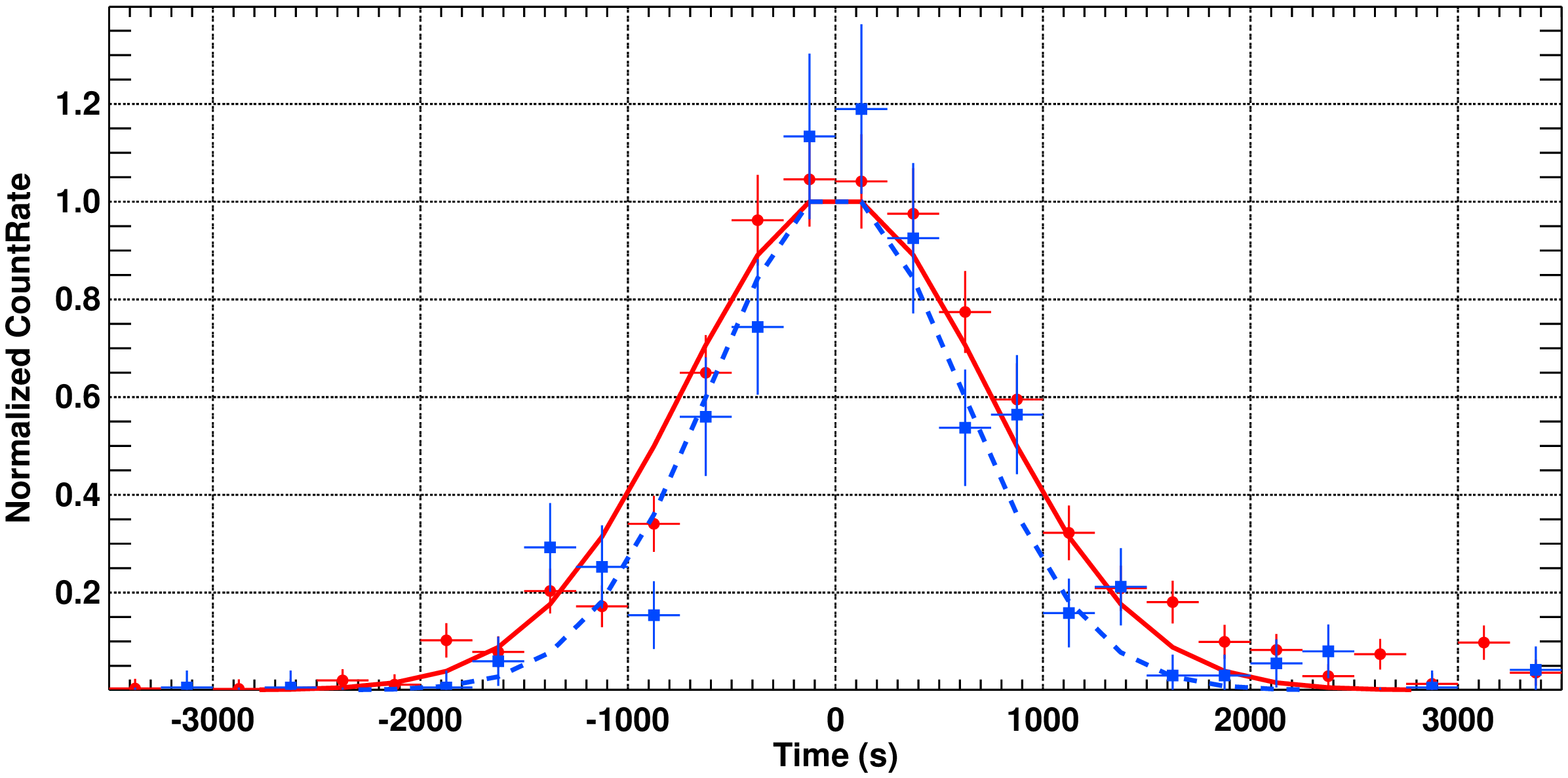}
\caption{ 
{\it (Upper panel)} The combination of the light curves of VB1, VB2 and VB3 
in the 2-4 keV (red squares), 4-6 keV (orange stars), 6-10 keV (blue squares), 
light curves are shifted by the center of their best-fit Gaussian profile, 
subtracted by the best-fit local underlying continuum, and normalised by 
the peak of their best-fit Gaussian. Time bins of 150~s are used. 
Flares are shorter in the hard band. 
{\it (Bottom panel)} Chandra composite light curve for 2-4.5 keV (red circles) 
and 6-9 keV (blue squares). All bright flares not significantly affected by pile-up 
(peak count rate $<0.1$~ph~s$^{-1}$) are considered here. The light curves 
are shifted with the same method described above. Time bins of 250~s are used.
Flares last longer in the 
soft band. }
\label{LCEneComp}
\end{figure}

The top panel of Fig. \ref{LCEne} suggests that \sgras's flares shorten 
with energy, typically lasting $\sim5~$\% less time in the hard band (6-10~keV) 
compared to the soft one (2-4~keV). The top panel of Fig. \ref{LCEneComp} 
shows the combination of all three very bright flares. With the solid red, 
dotted orange and dashed blue lines the light curves in the 2-4 keV, 
4-6 keV and 6-10 keV energy bands are shown. The light curves are 
shifted by the center of their best-fit Gaussian profile, subtracted by the 
best-fit local underlying continuum, and normalised by the peak of their 
best-fit Gaussian (see Tab. \ref{modLC}). 
The flare profile is tighter at higher energies
(Fig. \ref{LCEneComp}, Tab. \ref{modLC}). 
Indeed, the width of the Gaussian fitting the 
6-10~keV band appears to be significantly smaller (at $\sim4.4\sigma$ 
significance) by $\sim360$~s compared to the 2-4~keV band one 
(see Tab. \ref{modLC}).  
To test whether this is a common property of all X-ray flares or whether it 
is a peculiarity of very bright flares, we combined all \chandra\ bright 
and very bright flares (i.e. with fluence larger than 
$5\times10^{-9}$~erg~cm$^{-2}$; see Ponti et al. 2015a for the definition). 
To avoid flares significantly affected by pile up, we excluded the ones 
observed in either ACIS-I or ACIS-S with no subarray mode and reaching a 
block count rate equal or higher than 0.1~ph~s$^{-1}$ (see 
Ponti et al. 2015a for details). We also excluded the flares observed 
in ACIS-S 1/8 subarray mode and reaching a block count rate equal 
or higher than 0.8~ph~s$^{-1}$ (Ponti et al. 2015a). 
The bottom panel of Fig. \ref{LCEneComp} shows the combined 
\chandra\ flare light curve in the 2-4.5 and 6-9 keV energy bands with 
red circles and blue squares, respectively. We find that bright \chandra\ flares also
last longer in the soft energy band compared to the hard one, with 
a difference in $FWHM$ of $\sim300$~s (Fig. \ref{LCEneComp} and 
Tab. \ref{modLC}). 

No significant time shift with energy is apparent, with upper limits as tight 
as $\sim100-200$~s (see bottom panel of Fig. \ref{LCEne}). 

\begin{table}
\centering
\caption{Results of single Gaussian fitting to the 200 second binned flare light 
curves without background subtraction, but with {\tt epiclccorr} applied. 
Note that using the background subtracted light curves only changes these 
values very slightly. 
$^{1}$best-fit local continuum under the flare. 
$^{2}$delay time related to the 4-6 keV band. The errors are all 1 $\sigma$. 
The composite flare from \xmm\ is the combination of VB1, VB2 and VB3. 
The composite flare from \chandra\ is the combination of a set of unpiled-up 
flares observed by \chandra.}
\centering
\begin{tabular}{llllll}
\hline
Flare ID & Band & Contiuum$^{1}$ & $FWHM$ & Delay$^{2}$\\
	     & {\it keV} & \multicolumn{1}{c}{{\it $cts~s^{-1}$}} & \multicolumn{1}{l}{{\it s}} & \multicolumn{1}{l}{{\it s}}\\
\hline
VB1 &2 - 4    & $0.0446\pm0.0017$ & $1500\pm85$ & $+8\pm$49\\
        &4 - 6   & $0.0310\pm0.0014$ & $1550\pm57$ & 0\\
        &6 - 10 & $0.0128\pm0.0009$ & $1385\pm61$ & $+17\pm$42\\
\hline
VB2 &2 - 4   & $0.0485\pm0.0008$ & $1860\pm130$ & $-25\pm$73\\
	&4 - 6   & $0.0433\pm0.0008$ & $1570\pm80$ & 0\\	
       &6 - 10 & $0.0198\pm0.0005$ & $1610\pm100$ & $-16\pm$62\\
\hline
VB3 &2 - 4     & $0.0956\pm0.0016$ & $1840\pm153$ & $-48\pm$79\\
	&4 - 6     & $0.0893\pm0.0015$ & $1455\pm80$   & 0 \\	
        &6 - 10  & $0.0273\pm0.0008$ & $1280\pm75$   & $-100\pm$53\\
\hline
Composite     &2 - 4    & $0.0610\pm0.0007$ & $1813\pm68$ & --\\
(XMM)           &4 - 6    & $0.0559\pm0.0007$ & $1554\pm42$ & --\\	
        		 &6 - 10  & $0.0242\pm0.0004$ & $1450\pm47$ & --\\
\hline
Composite &2 - 4.5  & $7.9\pm0.3\times10^{-4}$ & $1730\pm56$ & --\\
(\chandra) &4.5 - 6  & $4.1\pm0.2\times10^{-4}$ & $1693\pm59$ & --\\	
       		&6 - 9 & $6.0\pm0.4\times10^{-4}$ & $1424\pm89$ & --\\
\hline
\end{tabular}
\label{modLC}
\end{table}

We conclude that, at present, the X-ray data of the bright flares 
are not in contradiction with a slow variation of $\gamma_{max}$. 

\section{Discussion}
\label{discussion}

Simultaneous \xmm, \nustar\ and \sinfoni\ observations of \sgras\ 
allowed us to determine, for the first time, the spectral shape 
and the evolution of the radiation of a very bright flare. 
This enabled us to pin down the radiative mechanism during bright 
flares of \sgras\ and its evolution during the flare. 
We can rule out that a simple power-law model, representing plain 
Synchrotron emission, can reproduce the flare emission. 

\subsection{TSSC} 

A TSSC model provides 
an acceptable fit to the data, from a statistical point of view. However: 
i) the fit is worse than the PLCool and PLCoolEv
models, despite the larger number of free parameters; 
ii) in this framework the observed spectral steepening 
($\Delta\Gamma=0.57\pm0.09$ at 1 $\sigma$) between NIR and X-ray
would be just a coincidence, moreover; 
iii) the best fit parameters appear implausible. 

Indeed, the best fit magnetic field appears unreasonably 
high ($B\sim10^4$), the source size unphysically small 
($R_F\sim10^{-2}-10^{-4}$~R$_S$) and the required source 
density about seven orders of magnitude higher than what is 
estimated to be present in the accretion flow around \sgras\ 
(Loeb \& Waxman 2007; Genzel et al. 2010).
As discussed in Dodds-Eden et al. (2009), these unreasonable best fit 
parameters are consequences of the assumptions intrinsic 
to the TSSC model considered here. In particular, to fit the 
soft X-ray emission via inverse Compton up-scattering of NIR or 
sub-mm radiation, the energies of the electrons involved in the flare 
is restricted to be lower than $\gamma_e<100$. On the other hand 
the requirement of the observed hard NIR slope constrains the magnetic 
field to be larger than $B>10^3$~G. Finally, the ratio of the synchrotron 
to inverse Compton luminosity requires that the size of the source has to
be $R_F<10^{-2}$~$R_S$. Therefore, it appears that this simplistic 
TSSC model, cannot adequately explain the flare emission. 

The observed spectral steepening ($\Delta\Gamma=0.57\pm0.09$, 
1 $\sigma$) between NIR and X-rays of the mean spectrum suggests that 
the radiative process during bright flares might be synchrotron 
radiation with a cooling break. Therefore, we explored in more  
details this scenario, instead of considering more complex 
synchrotron self Compton (SSC) models. 
Nonetheless, this does not rule out that more complex TSSC and 
non-thermal SSC models might be invoked to explain the X-ray 
radiation. 

\subsection{PLCool} 

We observed that the synchrotron model with a cooling break can 
reproduce both the mean spectrum and the evolution of the SED 
of \sgras\ during the entire duration of a very bright flare (apart from IR2). 
In particular, the observed spectral steepening of both the mean 
spectrum and of the emission at the X-ray peak (IR3) is a strong 
indication that synchrotron with cooling break might be the dominant 
radiative mechanism. 
In this simplistic model, the ``unknown" motor powers the continuous 
acceleration of energetic electrons with a power-law distribution. 
An important difference of the PLCool model compared to the TSSC model, 
is that the motor is assumed to accelerate electrons 
into a power-law distribution up to $\gamma_{max}\geq10^6$, 
therefore the synchrotron radiation is not limited to the NIR band, 
instead it extends to X-ray and higher energies. 

\subsubsection*{Limitations of the model}

We point out that, for simplicity, we reproduce the synchrotron 
emission with a simple broken power law model. We note that  
the cooling break typically occurs in the unobserved optical-UV band, 
therefore we can not currently constrain whether the cooling break 
is a sharp feature or it is significantly extended in energy. 
Indeed, we do not observe any significant curvature in either 
the NIR or X-ray band, however this has to be attributed to the 
small frequency windows sampled by our data. 
Therefore, for simplicity, we assume a sharp break, though realistic 
synchrotron models can be significantly broadened (by up to more 
than a decade in energy; Dibi et al. 2014). Indeed, it is beyond 
the scope of this paper to employ more complex synchrotron models. 
We also note that significantly different statistics characterise 
the time resolved spectra in the NIR and X-ray bands. Therefore, 
the broad band fit of the time resolved spectra are primarily driven 
by the NIR photon index. It is important to point out that at 
the beginning of the flare, the cooling break is observed to be located 
either within or very close to the NIR band. Therefore, should 
a broadened break be present at that time, it might potentially affect 
some of the model parameters. 
Future investigations will clarify the extent of this. 

\subsubsection*{Difficulties of the PLCool model}

As already briefly mentioned in \S \ref{EvPLCool}, we stress 
again here that the most difficult problem of the PLCool 
model is related to the best fit energy of the cooling break. 
Indeed, both for IR1 and IR2 the break is observed within the NIR 
band (see Tab. \ref{TabSEDev}). During IR1 this result is driven 
by the combined upper limit on the X-ray emission and by the high 
flux and flat photon index in the \sinfoni\ band, inducing a cooling 
break suspiciously located at energies just higher than $\sim0.6$~eV, 
the upper bound of the \sinfoni\ spectrum. Similar results holds 
during IR2. Indeed, at that time, the X-ray flare had already started,
alleviating the problem, however the NIR band showed a slightly flatter 
power-law ($\Gamma_{NIR}=1.59\pm0.2$; Tab. \ref{IRtime}), therefore 
the cooling break 
was observed to again be placed at the upper bound of the \sinfoni\ spectrum
(moreover the steep X-ray slope is not completely reproduced). 
We consider a rather unlikely possibility that, by chance, the cooling 
break occurred twice within the narrow NIR band. 

The biggest pitfall that the PLCool model has to overcome 
is the explanation of the early phases of the VB3 flare. Indeed, 
at the basis of the PLCool model there is the assumption that 
the NIR and X-ray emissions are tied by a broken power law. 
Therefore, within this framework one would predict that the NIR 
to X-ray emission are strictly related and they follow each other. 
The only deviation to this "rule" could be generated by the possible 
delay of the NIR radiation associated with the longer NIR synchrotron 
cooling time. Therefore, it is expected that the X-ray emission either 
rises before or at the same time as the NIR one. One possible way 
out (that has been considered in the past to explain the delayed 
X-ray emission in the early phases of the very bright flares) 
was to assume that the early NIR emission had a very steep slope. 
However, we can now rule out that this is happening during VB3. 
Indeed, during IR1 bright NIR emission, with a flat slope 
($\Gamma_{NIR}=1.48\pm0.2$; Tab. \ref{IRtime}), is observed at 
the same time of tight upper limits to the X-ray emission. 
On the contrary, no prominent X-ray radiation is observed either 
during or before IR1, with upper limits in the 3-10 keV band 
of $F_{3-10~keV}<2.3\times10^{-12}$~erg~cm$^{-2}$~s$^{-1}$.

The third difficulty of the PLCool model is to properly reproduce 
the IR2 spectrum. A fit with the BPL model shows that the difference 
of the X-ray to NIR photon index is significantly higher 
$\Delta\Gamma=1.8\pm0.4$ than the one expected by the cooling 
break model $\Delta\Gamma=0.5$ 
(Tab. \ref{TabSEDev}). This resulted in a poor fit of IR2 by the PLCool 
model (indeed an F-test suggests that the BPL model provided 
a significantly better description of the IR2 spectrum, at 
$>99$~\% confidence). 

\subsection{PLCoolEv}
\label{SecDiscPLCoolEv}

We then relaxed the requirement that $\gamma_{max}$ has 
to be $>10^6$ at all times. We postulated that, $\gamma_{max}$ 
increases slowly with time (e.g. many times the Alfv\'en speed 
crossing time of a source of a size of few Schwarzschild radii), 
generating a cut-off that gradually moves to higher energies, 
eventually transiting through the X-ray band and producing a
bright X-ray radiation with a delay compared to the start of the 
NIR flare. Regardless of the behaviour of $\gamma_{max}$ 
at the end of the flare, the PLCoolEv (as well as PLCool) model 
predicts a delay of a few hundred seconds of the NIR radiation, 
compared to the X-ray emission (\S \ref{Evgamma}), in agreement 
with a longer duration of the NIR flare.  

\

\subsubsection*{Limitations of the model}

For simplicity, we assumed an exponential drop of the high energy 
cut off in the synchrotron spectrum, with a shape such that the e-folding 
energy is equal to the cut off energy. We note that the high energy 
cut off is detected in the observed band only once, during IR2. 
During this interval, the X-ray slope is steeper ($\Gamma=3.2\pm0.4$) 
than the simultaneous NIR one, however the statistics is not enough 
to discriminate its detailed shape. For example, we could not distinguish 
either between an exponential or a sub-exponential, or we could not 
constrain the broadness of the cut off. 
Therefore, should a broadened break be present at that time, 
it might potentially affect some of the model parameters. For instance, 
broader cutoffs in IR2 and IR4 might allow for a cooling break at higher 
energy, implying a weaker magnetic field. Future investigations will clarify 
the extent of this.

\subsubsection{Comparison of the PLCoolEv model to the data}

The PLCoolEv model provides an excellent description of the mean 
spectrum of VB3 and of its evolution over time. 
In fact, it naturally explains the periods during which bright 
and flat-spectrum NIR radiation is observed, simultaneous with 
no X-ray emission (e.g. IR1). In particular, the passage of the cut-off 
within the X-ray band generates: i) shorter flare durations at higher 
energies; ii) right at the start of the X-ray flare steeper X-ray 
spectra than expected by the cooling break (e.g. IR2) and; 
iii) possibly steeper spectra in the flanks of the X-ray flare than 
at the peak, such as observed. 

We also measured a significant evolution of the cooling break 
during the flare. Under the assumption that the escape time 
remains constant, this suggests that the strength of the magnetic 
field (typically of several tens of Gauss) lowers to values of few Gauss 
during the peak of bright flares, to return to high values after that. 
For a thermal distribution of electrons with temperature $\theta_E$, 
the Synchrotron luminosity is proportional to the square of the magnetic 
field ($L_{Synch}\propto N\theta^2_E B^2$), therefore the drop of 
the magnetic field strength at the flare peak would appear contradictory. 
However, in this scenario, the large Synchrotron luminosity is provided 
by the vast increase in the energy of the accelerated particles. 
It is likely that the acceleration mechanism is powered by the magnetic 
field, that therefore gradually reduces its strength during the flare 
(Dodds-Eden et al. 2009; 2010; 2011). 
Indeed, a similar process is at work in magnetic reconnection that 
is a fundamental process of plasmas in which magnetic energy is 
converted into particle acceleration through magnetic field 
rearrangement and relaxation (Begelman 1998; Lyubarski 2005; 
Zweibel \& Yamada 2009; Sironi et al. 2014; 2016). 
The observed drop of the magnetic field, right at the peak of the 
flare, is in line with the predictions of magnetic reconnection models. 

We conclude that the data are consistent with such an evolution 
of the magnetic field. 

\subsubsection{Constraints on energy power and source size}
\label{Ssize}

{\it Is the energy stored in the magnetic field enough to power the flare?} 
We estimated the total energy emitted during the VB3 flare by considering 
that the NIR and X-ray luminosity was at a level of $\sim10$~mJy and 
$\sim2\times10^{35}$ erg s$^{-1}$ for about $\sim3.8\times10^3$~s and 
$\sim1.8\times10^3$~s, respectively, resulting in a total emitted energy of 
$\sim8\times10^{38}$~erg. If we discharge the magnetic energy within 
a spherical region with $\sim1.5$~$R_S$ radius, bringing its magnetic 
field from $B\sim30-40$~G to $\sim5$~G, then about 
$\sim8-15\times10^{38}$~erg are produced. This appears to be 
enough energy to power the VB3 flare. Moreover, this suggests that the 
source of VB3 had a size $\geq1.5$~$R_S$.  

\section{Conclusions}
\label{conclusions}

\begin{itemize}

\item{} The mean X-ray photon index during the very bright 
flare VB3 is significantly steeper ($\Gamma_X=2.27\pm0.12$) 
than the simultaneous ($\Gamma_{NIR}=1.7\pm0.1$) NIR one, 
excluding that the radiative process can be described by a simple 
power-law. In particular, the observed steepening 
($\Delta\Gamma=0.57\pm0.09$ at 1 $\sigma$) 
is consistent with what is expected by Synchrotron emission with 
a cooling break ($\Delta\Gamma=0.5$). 

\item{} We observe bright $F_{2.2\mu m}=8.9\pm0.1$~mJy and 
hard NIR ($\Gamma_{NIR}=1.48\pm0.23$) emission about $\sim10^3$~s 
before the start of the X-ray flare. We also observe very steep 
X-ray emission ($\Gamma_X=3.2\pm0.4$) at the start of the X-ray flare, 
while the contemporaneous NIR photon index was $\Gamma_{NIR}=1.4\pm0.2$. 
These results strongly support a scenario where the synchrotron 
emitting electron power-law distribution has a cut-off ($\gamma_{max}$) 
that is {\it slowly} evolving with time, therefore inducing an evolving 
high energy cut-off in the spectrum. 

\item{} The data are consistent with an evolution of the magnetic field 
strength during the flare (under the assumption of a constant escape time). 
Large magnetic field amplitudes ($B=30\pm8$~G) 
are observed at the start of the X-ray flare. The magnetic field strength 
drops to $B=4.8\pm1.7$~G, at the peak of the X-ray flare, a variation 
of a factor of $>6$ in less than $\sim650$~s. It then increases again 
in the decreasing flank of the flare ($B=14.3^{+12.3}_{-7.0}$~G). 
This is consistent with a scenario where the process that 
accelerates the electrons producing the Synchrotron emission 
is tapping energy from the magnetic field (such as, e.g. in magnetic 
reconnection). 

\item{} From the total emitted energy and the variation of the 
magnetic field, we estimated that the source size of the VB3 flare 
has to be larger than $\geq1.5$~$R_S$, if powered 
by magnetic reconnection. 

\item{} We observe hints for steeper, by roughly $\Delta\Gamma=0.3$, 
X-ray spectra during the rise and the decay of an X-ray flare, compared
to the values at peak. This indicates that, despite the fact that the 
photon index is similar between different X-ray flares, there might be 
significant spectral evolution during each X-ray flare. This is an 
expectation of the PLCoolEv model. 

\item{} Bright and very bright \xmm\ and \chandra\ flares typically last 
significantly ($\sim4.4\sigma$ significance) longer, by $\sim300$~s at 
soft X-ray energies, compared to harder ones (2-4 and 6-10~keV, respectively).
This trend appears to join smoothly to the longer duration typically observed 
in the NIR band. Again, this is most probably the product of the evolution 
of $\gamma_{max}$.

\item{} The three very bright flares, caught by \xmm\ so far, have very similar 
light curves and spectral properties, indicating an analogous physical origin. 
This suggests that the results of this study on VB3 could be universal to bright 
and very bright flares. 

\item{} The best fit column density of neutral absorbing material observed 
during the X-ray spectra of the very bright flares of \sgras\ is constant and 
it is consistent with the values observed in nearby sources. 
Indeed, the three bright transients within $d_{\rm pro}<1.5^\prime$ from 
\sgras\ (\sgr, \ssv\ and the foreground component towards \axj) show neutral 
absorption column densities consistent with the value of \sgras\ 
(Coti-Zelati et al. 2015; Ponti et al. 2016a; 2016). This suggests that 
the neutral absorption towards \sgras\ has an ISM origin. 

\item{} Synchrotron self Compton models can statistically reproduce the flare 
emission and its evolution. On the other hand, they imply unrealistic parameters. 
In such a scenario it would 
be an unlikely coincidence that the NIR photon index is flatter than the X-ray 
one by $\Delta\Gamma=0.5$. Moreover, the evolution of the density, 
source radius and magnetic field before, during and after the very bright flare 
appears improbable. 

\end{itemize}
 
\section*{Acknowledgments}

The authors wish to thank Jan-Uwe Ness, Ignacio de la Calle, Karl Foster 
and the rest of the \xmm\ and \nustar\ scheduling teams for the enormous 
support that made this multi-wavelength campaign possible, as well as the 
referee for the careful reading of the paper. GP thanks Lorenzo Sironi, 
Hendrik J. van Eerten, Michi Baub\"{o}ck and Francesco Coti-Zelati, 
for useful discussion. RT and AG acknowledge support from CNES. 
This research has made use both of data obtained with \xmm, an ESA 
science mission with instruments and contributions directly funded by ESA 
Member States and NASA, and on data obtained from the Chandra Data Archive.
The GC \xmm\ monitoring project is supported by the Bundesministerium 
f\"{u}r Wirtschaft und Technologie/Deutsches Zentrum f\"{u}r Luft- und Raumfahrt 
(BMWI/DLR, FKZ 50 OR 1408 and FKZ 50 OR 1604) and the Max Planck Society. 

\appendix

\section{Further details on \xmm\ data reduction}

All X-ray observations considered here have been accumulated with the EPIC-pn 
and EPIC-MOS cameras in Full Frame mode with the medium filter applied
(apart from {\sc ObsID 0111350301} which has the pn camera with the thick filter;
see Ponti et al. 2015a,b). \sgras's flares in Tab. \ref{obsid} 
occurred during periods of negligible soft proton flare activity, therefore no cut 
was performed during those flares. On the other hand, significant soft proton 
flares are detected during quiescent emission. We removed these periods 
of enhanced background activity by cutting all intervals with more 
than 0.25 ph~s$^{-1}$ in the background light curve (integrated over the 
10-15~keV energy band, with 20~s time bins and extracted from a $3^{\prime}$
radius). 
We selected only single and double events and we used (FLAG == 0) and 
either (\#XMMEA\_EP) or (\#XMMEA\_EM) for EPIC-pn or MOS, respectively. 
We applied the {\sc sas} task {\sc lccorr} to the \xmm\ light curves. 

We note that during obsID: 0743630201, 0743630301 and 0743630501 
\sgras's flux is contaminated by the X-ray emission from the magnetar 
SGR~J1745-2900, located at only $\sim2.4^{\prime\prime}$ from \sgras\ 
(Degenaar et al. 2013; Mori et al.\ 2013; Rea et al.\ 2013). 
During these observations the magnetar's flux was 
$F_{\rm 1-10~keV}\sim3\times10^{-12}$~erg~cm$^{-2}$~s$^{-1}$ (see Coti Zelati 
et al. 2015; for the details of the decay curve), therefore allowing an adequate 
characterisation of the bright flares. 

$\ddagger$ We report in Tab. \ref{obsid} the same time systems 
as used by \xmm\ (see \S~6.1.4 in \xmm\ Users Handbook, 
https:\/\/heasarc.gsfc.nasa.gov\/docs\/xmm\/uhb\/reftime.html). 
The reference or zero time has been defined as: 1998-01-01T00:00:00.00 TT 
= 1997-12-31T23:58:56.816 UTC. The conversion from TT to UTC at the 
reference date is TT = UTC + 63.184 s and can be derived from The 
Astronomical Almanac for other dates.

\section{Spectral distortions introduced by dust scattering}
\label{SecDSH}

As already discussed in \S \ref{IntroDSH}, dust scattering does 
severely distort the source spectrum and if its effects are not 
properly taken into account, it could significantly bias the results. 
To provide the reader with a better understanding of the extent of 
this effect on the various best fit parameters, we present in 
Tab. \ref{TabDSH} the best fit results of the mean X-ray spectrum 
of VB3 after the application of: i) no correction; ii) corrections with 
the {\sc dust} model and; iii) correction with the {\sc fgcdust} model. 
For these fits we left the column density of neutral absorbing material 
free to vary. 
When we applied the {\sc dust} model, we tied the dust scattering 
optical depth $\tau$ to the fitted column density so that $\tau=0.324 
(N_H/10^{22}~cm^{-2})$, following Nowak et al. (2012). 
We also assumed a ratio of 10 between the size of the halo at 1 keV and 
the extraction region. 
\begin{table}
\centering
\caption{Best fit parameters, once the mean X-ray spectrum of VB3 
is corrected with different dust scattering models. Column densities are 
in $10^{23}$~cm$^{-2}$ units. Fluxes are in $10^{-12}$~erg~s$^{-1}$~cm$^{-2}$
units, integrated over the $3-10$~keV band and are absorbed, 
but corrected for the effect of dust scattering. The \xmm\ spectrum 
is fitted with an absorbed power-law modified by the dust scattering. }
\centering
\begin{tabular}{l c c c}
\hline
                   & No dust                       & {\sc dust}                & {\sc fgcdust}              \\
\hline
$N_H$         & $1.8\pm0.3$              & $1.3\pm0.2$            & $1.6\pm0.3$            \\
$\Gamma$  & $2.0\pm0.3$              & $2.1\pm0.3$            & $2.2\pm0.3$             \\
$F_{3-10}$  & $7.4^{+10.8}_{-3.6}$ & $8.3^{+7.2}_{-3.7}$ & $8.9^{+9.0}_{-4.4}$  \\
$\chi^2/dof$ & $125.8/116$              & $124.2/116$            & $126.9/116$             \\
\hline
\end{tabular}
\label{TabDSH}
\end{table}

The best fit results in Tab. \ref{TabDSH} show that the inclusion of the 
dust scattering model allows us to recover a steeper and brighter 
source spectrum. 
Indeed, the main effect of dust scattering is to remove flux from the line of sight 
and to spread it in the halo, that is typically partly lost because of the small 
source extraction region. Moreover, the probability of dust scattering is 
higher at low energy, producing a deficiency of low energy photons in the 
observed spectrum (that is generally reproduced by a higher column density 
of absorbing material and flatter spectra). 
Therefore, once the correction for dust scattering is introduced, we observe 
that the flux increases by $\sim20$~\%, the column density of neutral 
material is lower and the spectrum steepens. In particular, we observe 
the photon index to steepen by $\Delta\Gamma\sim0.2$. It is important 
to note that different dust models lead to photon indexes that differ 
by $\Delta\Gamma=0.1$ (see Tab. \ref{TabDSH}). 

For this reason, we performed again all the analyses in the paper, correcting 
the effects of dust scattering with the {\sc dust} model. 
The fit of the mean VB3 X-ray spectrum results in a $\Gamma_X=2.16\pm0.14$, 
that once compared to the NIR slope $\Gamma_{NIR}=1.7\pm0.1$, 
gives $\Delta\Gamma=0.46\pm0.17$ ($\pm0.10$ at 1 $\sigma$), which is 
completely consistent with the PLCool model ($\Delta\Gamma=0.5$). 
The same applies also to the IR3 interval. Indeed, the X-ray photon index 
during IR3 is $\Gamma_{X}=2.5\pm0.3$, therefore steeper 
by $\Delta\Gamma=0.6\pm0.3$, compared to the simultaneous NIR 
measurement ($\Gamma_{NIR}=1.9^{+0.1}_{-0.2}$). Therefore, it is in 
this case also consistent with the conclusions of this work. 

\begin{figure}
\includegraphics[height=0.49\textwidth,angle=90]{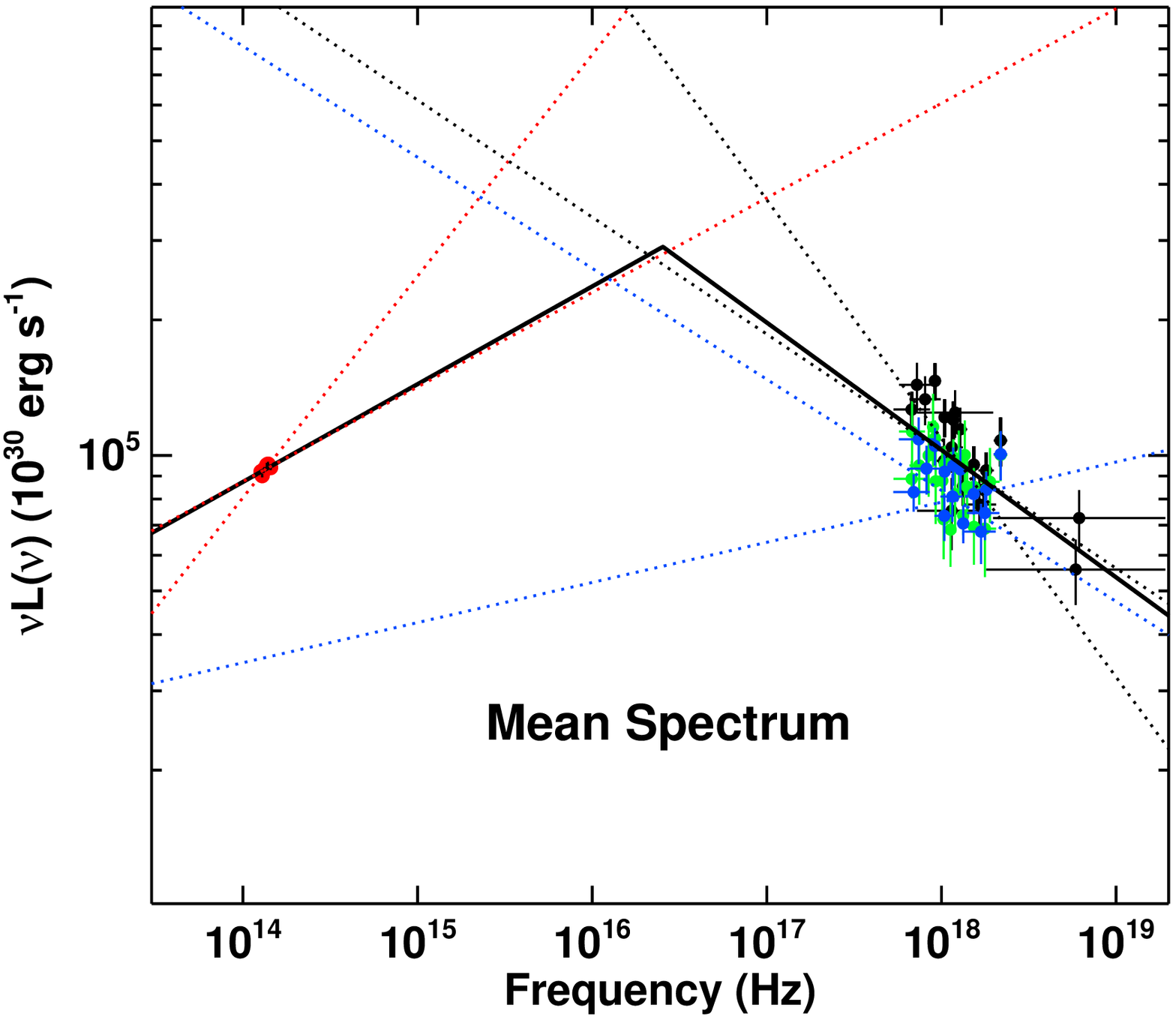}
\includegraphics[height=0.49\textwidth,angle=90]{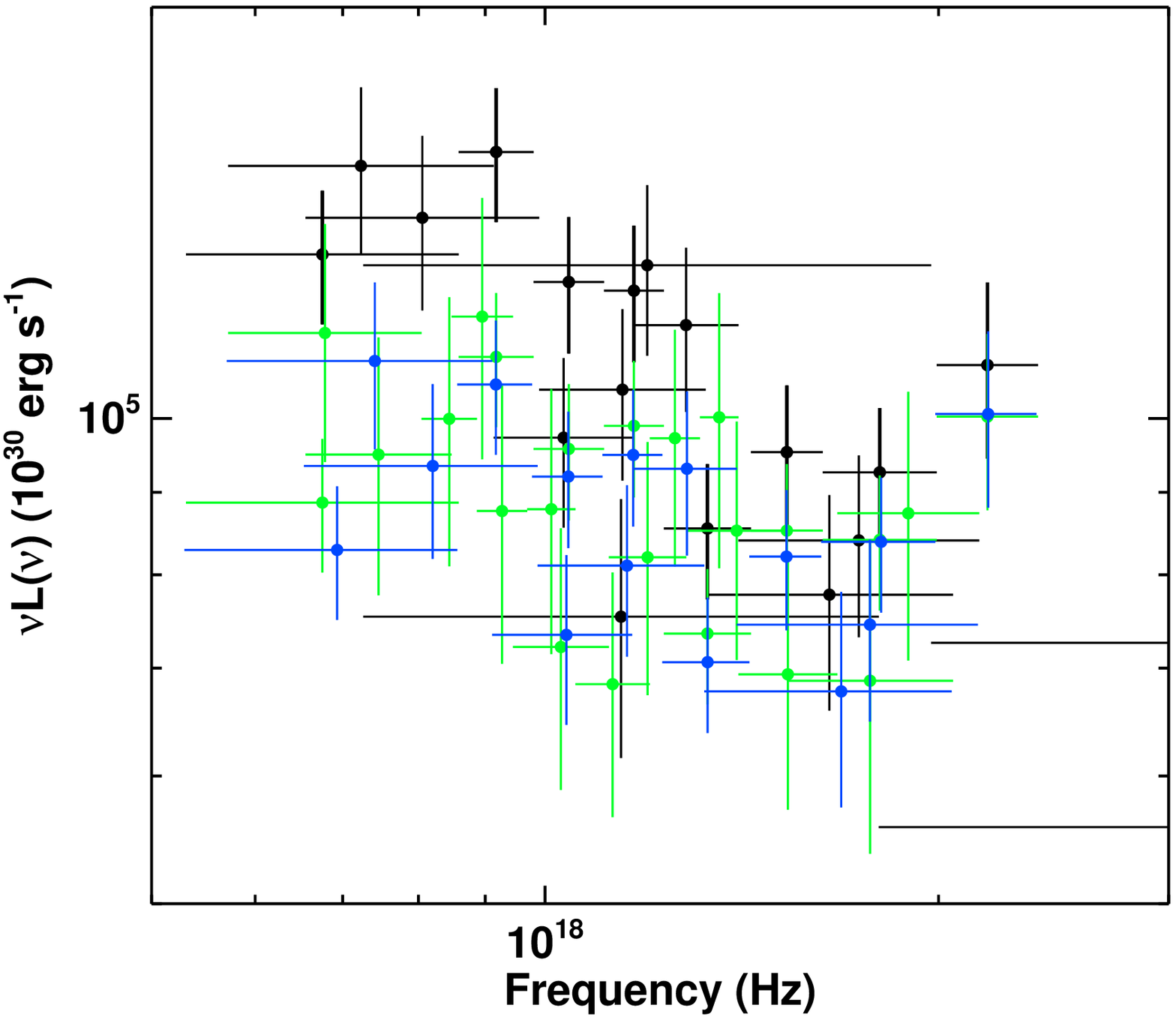}
\caption{{\it (Top panel)} 
The red points show the mean NIR (\sinfoni) during the VB3 flare, 
such as in Fig. \ref{PLCool}. The blue points show the corresponding 
X-ray (\xmm\ and \nustar) spectra, when no dust correction is applied. 
The green and black points show the X-ray spectra corrected for the 
effects of dust scattering according to the {\sc dust} and {\sc fgcdust} 
models, respectively. The stronger the correction, the brighter and 
steeper becomes the X-ray spectrum (see text for details). 
The red, blue and black dotted lines show the uncertainty on the NIR slope, 
on the X-ray slope when no correction for dust is applied and once the 
data are corrected using the {\sc fcgdust} model, respectively. The black line shows 
the best fit PLCool model (where the X-ray and NIR slopes are tied). 
All data and models and data are de-absorbed. 
{\it (Bottom panel)} Enlargement of the top panel into the X-ray band, 
to allow a better elucidation of the extent of the effect. }
\label{FigDSH}
\end{figure}

The blue, green, and black data in Fig. \ref{FigDSH} show the \xmm\ de-absorbed 
mean spectrum of VB3 (see also Fig. \ref{PLCool}) after correcting the observed 
spectrum for the effects of dust scattering with the {\sc fgcdust}, {\sc dust} 
models and after applying no correction, respectively. The blue dotted lines 
show the uncertainties in the determination of the X-ray slope if no correction 
for dust scattering is applied. Even in this case, the difference in X-ray and 
NIR spectral shapes are consistent with the predictions of the synchrotron 
model with cooling break. 

\end{document}